RUSSIAN ACADEMY OF SCIENCES
PETERSBURG NUCLEAR PHYSICS INSTITUTE

V. G. Gorshkov, A. M. Makarieva, B.-L. Li*

# Comprehending environmental and economic sustainability: Comparative analysis of stability principles in the biosphere and free market economy

*Department of Botany and Plant Sciences,
University of California, Riverside, USA*

**Сравнительный анализ экологической устойчивости биомассы биосферы и экономической устойчивости товаров свободного рынка**

В. Г. Горшков, А. М. Макарьева, Б.-Л. Ли

**Аннотация**

С использованием формализма потенциальной функции Ляпунова показано, что принципы поддержания устойчивости биомассы в экосистеме и занятости в экономике математически подобны. Экосистема имеет устойчивое и неустойчивое стационарные состояния с большой (лес) и малой (луг) биомассой, соответственно. В экономике существует устойчивое стационарное состояние с высокой занятостью при массовом производстве традиционных товаров с ценой, равной себестоимости, и неустойчивое стационарное состояние производства новых товаров технологического прогресса с низкой занятостью. Для экономики дополнительно описано второе устойчивое состояние производства сырья и энергии. В этом состоянии цивилизация платит 10% мирового ВНП за энергию, производимую порядка 0.2% мирового работающего населения и продаваемую по ценам, в 40 раз превышающим себестоимость. Показано, право собственности на источники энергии эквивалентно приравниванию измеримых величин разных размерностей (запасов и потоков) и приводит к эффективному нарушению законов сохранения вещества и энергии.

**Abstract**

Using the formalism of Lyapunov potential function it is shown that the stability principles for biomass in the ecosystem and for employment in economics are mathematically similar. The ecosystem is found to have a stable and an unstable stationary state with high (forest) and low (grasslands) biomass, respectively. In economics, there is a stable stationary state with high employment, which corresponds to mass production of conventional goods sold at low cost price, and an unstable stationary state with lower employment, which corresponds to production of novel goods appearing in the course of technological progress. An additional stable stationary state is described for economics, the one corresponding to very low employment in production of life essentials such as energy and raw materials. In this state the civilization currently pays 10% of global GDP for energy produced by a negligible minority of the working population (currently ~0.2%) and sold at prices greatly exceeding the cost price by 40 times. It is shown that economic ownership over energy sources is equivalent to equating measurable variables of different dimensions (stores and fluxes), which leads to effective violation of the laws of energy and matter conservation.

## 1. Introduction

Biomass of any type exists in the biosphere in the state of dynamic ecological equilibrium sustained by the balance of synthesis and decomposition. Biomass is synthesized by living organisms, i.e. by living biomass itself. At zero biomass the rate of its synthesis is zero. It grows with increasing biomass of the synthesizing organisms. However, biomass synthesis is ultimately limited by the incoming flux of solar radiation. When all the solar radiation available for synthesis is claimed by the synthesizing organisms, the rate of biomass synthesis reaches its maximum value, which is biomass-independent.

In contrast, decomposition of organic matter by living organisms does not depend on solar radiation. There are no physical limits to biological decomposition, which could, in principle, occur at any rate exceeding the rate of synthesis. If the rate of decomposition permanently exceeded the rate of synthesis, all biomass would have been destroyed and life on Earth ceased. In this paper, using the formalism of Lyapunov potential function, we determine the quantitative conditions accounting for the stability of life and explore why these conditions are apparently fulfilled in the biosphere (Section 2).

The amounts of goods and services instantly available in the market are, similarly to biomass in the biosphere, maintained in the state of dynamic equilibrium by the processes of their production and consumption, i.e. by the balance of offer and demand. We demonstrate that the equations for biomass change in the biosphere are mathematically equivalent to the equations for employment in the output of goods and services in economics. In economics a universal monetary indicator (market price) of the amounts of goods and services available in the market is employed. Using Lyapunov function we determine conditions for the stability of market prices on goods and services (Section 3). It is found that there is a non-zero stable equilibrium value of market price, which can be meaningfully interpreted as equal to the cost price, the latter numerically defined in terms of labor costs associated with the production of goods and services.

For the novel expensive goods, i.e. goods that appear as products of technological progress, market prices are found to be unstable. Such goods are characterized by an unsaturated demand and, at the moment of their appearance in the market, have a market price significantly in excess of their cost price. Expansion of the output of such goods with an increase in the number of people involved in the corresponding production processes leads to an increase of the amount of goods available in the market and fall of their high initial price. With market price of the novel expensive goods diminishing towards their cost price and with gradual saturation of the demand, the goods enter the category of conventional goods, i.e. goods that, at any time of civilization development, make up most of the consumer basket and have a stable market price (Section 4).

Energy and raw materials are used as the indispensable foundation for production of all goods in the civilization and, as such, represent a very special phenomenon in economics, being fundamentally distinct from any other economic realities. If these distinctions are ignored (as is the case in modern economic theory) and energy and raw materials are considered as free market commodities, then, as follows from analysis of Lyapunov function, their market price, unlike that of novel expensive goods, remains stable at its maximum value. By analyzing the world employment data for the energy sector, it is estimated that the average cost price of energy is several dozens of times lower than its modern market price (Section 5).

Unlike with novel expensive goods, the maximum market price of raw materials and energy can never be reduced, because consumption of raw materials and energy is, at any time, saturated and cannot be dramatically increased. Neither can production of raw materials be increased by rising employment in the corresponding economics sectors. Therefore, market price of raw materials and energy remains fixed at its maximum value. This value is the maximum price that the world economy can still afford to pay without self-destruction, currently spending on energy about one tenth of global gross domestic product. Therefore, with increasing global gross domestic product market prices of energy can only grow (Section 6).

The dimensional analysis of stores and fluxes in economics proves that privatization of sources of raw materials and energy is equivalent to equating measurable variables of different dimensions. This violates the laws of matter and energy conservation in economics (Section 7), thus decelerating technological progress and economic growth, destabilizing modern civilization and barring solutions of global environmental problems (Section 8).

## 2. Stability of biomass in the biosphere

Temporal changes of the amount $M$ (dimension kg) of any type of biomass in the biosphere are dictated by the law of matter conservation (see, e.g., Lotka, 1925):

$$\frac{dM}{dt} = P^{+}(M) - P^{-}(M)\,, \qquad (1)$$

where $P^{+}(M)$ and $P^{-}(M)$ are the biomass-dependent production (synthesis) and consumption (decomposition) of this type of biomass (dimension kg · year$^{-1}$), respectively. All types of biomass production in the biosphere are determined by an external flux of energy, mainly solar radiation for photosynthesis and a million-fold smaller flux of concentrated geothermal energy for chemosynthesis.

In mature ecosystems production reaches its maximum value $P^{+}(M) = P^{+}$, which is determined by the efficiency of light capture by plants, and ceases to depend on biomass $M$. Biomass consumption in this case grows proportionally to the accumulating biomass $M$, which is close to some optimal stationary value and exceeds a certain value $M_r$, below which ecosystem degradation is possible. The value of $M_r$ is determined below. We can thus re-write Eq. (1) as

$$\frac{dM}{dt} = P^{+} - \frac{M}{T^{-}} = \frac{M_s - M}{T^{-}}\,, \quad M \geq M_r\,, \quad M_s \equiv P^{+}T^{-}\,. \qquad (2)$$

Here $T^{-}$ is the biomass turnover time via its consumption. Generally, any dependence $P^{-}(M)$ in (1) can be written as $P^{-}(M) = M/T^{-}(x)$, where $T^{-}(x) \equiv T^{-} f(x)$, $x \equiv (M_s - M)/M_s$ and $T^{-}$ is a constant magnitude independent of $x$. In the vicinity of the stationary point $x = 0$ we have $f(0) = 1$ and Eq. (2) is exact. One can expand $f(x)$ in a Taylor series about $x = 0$ and take into account the non-linear correctional $x$ terms (see, e.g., Lotka, 1925). We will assume that Eq. (2) at constant $T^{-}$ is true for any $x$ neglecting all the non-linear terms. Such a consideration is precise for the stationary point in (2) and insignificantly distorted when reasonably far from it, which does not affect any further conclusions.

Eq. (2) describes a mature ecosystem, where biomass $M$ randomly fluctuates about its optimal value $M_s \equiv P^{+}T^{-}$. At fixed production $P^{+}$ stationary biomass $M_s$ is low at short, and high at long, time $T^{-}$ of biomass turnover via its consumption. For example, biomass of epilithic lichens forming a thin cover on the rock surfaces is very small due to a short turnover time of biomass via its consumption by the lichen fungi. An opposite example is the large stationary biomass of forest wood, which is well protected against decomposition and has a very long turnover time of its consumption by bacteria and fungi of the forest community.

During recovery after natural disturbances in the process of succession, production grows proportionally to the accumulating biomass. For ecosystem production to grow, consumption of biomass $P^-(M)$ in (1) during succession should, obviously, grow with increasing biomass much less rapidly than does biomass production, i.e. it should remain relatively constant. Let us denote this biomass-independent value of consumption as $P^-$. In this case Eq. (1) takes the following form:

$$\frac{dM}{dt} = \frac{M}{T^+} - P^- = \frac{M - M_u}{T^+}, \quad M < M_r, \quad M_u \equiv P^- T^+. \tag{3}$$

As in Eq. (1), in Eq. (3) we neglect possible non-linear terms correctional to the main dependencies of production and consumption on biomass captured in Eq. (3); this does not affect any subsequent conclusions.

Stability of stationary states can be studied with use of Lyapunov potential function $U$ defined as

$$\frac{dM}{dt} = -\frac{dU}{dM}. \tag{4}$$

Stationary points correspond to the extremes of $U$, which represent stable minima at $d^2U/dM^2 > 0$ and unstable maxima at $d^2U/dM^2 < 0$.

The value of $M_r$ (2), (3) can be obtained by joining Eqs. (2) and (3) at $M = M_r$, i.e. by equating the right-hand parts of these equations (and the first derivatives of potential function $U(M)$ (4)), in this point. We thus obtain

$$M_r = \frac{M_s T^+ + M_u T^-}{T^+ + T^-}, \quad M_s \equiv P^+ T^-, \quad M_u \equiv P^- T^+. \tag{5}$$

We now integrate Eqs. (4) with use of Eqs. (2) and (3), setting the constant of integration in (3) equal to zero, and choosing the constant of integration $r$ in (2) from the continuity condition for function $U$ at $M = M_r$. Now, using Eq. (5) for $M_r$, we obtain the following expressions for the potential function, which describes both solutions, (2) and (3), Fig. 1:

$$U(M) = \begin{cases} -\dfrac{MM_s}{T^-}\left(1 - \dfrac{1}{2}\dfrac{M}{M_s}\right) + r, & M \geq M_r, \\ \dfrac{MM_u}{T^+}\left(1 - \dfrac{1}{2}\dfrac{M}{M_u}\right), & M \leq M_r, \end{cases} \tag{6}$$

$$r \equiv M_r^2 \frac{(T^+ + T^-)}{2T^+T^-}. \tag{7}$$

In Fig. 1 one can see Lyapunov function $U(M)$ (6) at different values of the fundamental parameter $\alpha \equiv (P^-T^+)/(P^+T^-) \equiv M_u/M_s$. At $\alpha < 1$, Fig. 1a, $U(M)$ has a stable minimum at $M = M_s$ and an unstable maximum at $M = M_u$. With $\alpha$ tending to unity ($M_u \to M_s$), the maximum approaches the minimum. At $\alpha = 1$ ($M_u = M_s$) the maximum and minimum coincide; the two extremes form a single unstable stationary point of inflection, Fig. 1b. At $\alpha > 1$ ($M_u > M_s$) function $U(M)$ has no extremes, Fig. 1b. In this case there are no stationary, time-independent states of the ecosystem.

Thus, for biomass to be stable, the ecosystem must have two possible states: one with fixed production $P^+$, maximum stationary biomass and decomposition increasing with growing biomass (forest) and another with fixed decomposition $P^-$, minimum stationary biomass and production $P^-$ increasing with growing biomass (e.g., grassland).

Stability of biomass and, hence, of life in the biosphere is only possible at $\alpha << 1$. In this case the value of biomass can fluctuate randomly within the right potential pit, with the amplitude of these fluctuations never reaching up to the unstable maximum. If going over the unstable potential barrier of the maximum, the value of biomass will then drop unimpeded down to zero, which will mean end of life. The fundamental condition $\alpha <<$ 1 can be interpreted as follows. At $T^+/T^- \sim 1$ it implies $P^-/P^+ << 1$, which means that when production and consumption change at similar rates with increasing biomass, see (2) and (3), the maximum, biomass-independent value of production, $P^+$ in (2), is much large than the biomass-independent value of consumption, $P^-$ in (3). At $P^-/P^+ \sim 1$, i.e. when these mass-independent values are similar, $\alpha << 1$ implies $T^+/T^- << 1$, which means that $P^+(M) \equiv M/T^+$ (3) grows with biomass $M$ at $M < M_r$ much more rapidly than consumption $P^-(M) \equiv M/T^-$ (2) grows with biomass $M$ at $M > M_r$. When $P^+$ and $P^-$ and $T^+$ and $T^-$ come close, and there is a non-zero probability of going over the potential barrier, nothing can prevent the extinction of life in the biosphere. The fundamental parameters $P^+$, $P^-$, $T^+$ and $T^-$ can be experimentally determined for any particular biomass type by studying stable mature ecosystems as well as the dynamics of ecosystem recovery after disturbances. On land stable ecosystems are represented by forests featuring maximum biomass. Savannas, steppes and prairies are unstable with respect to transition to either forest or desert (Tutin et al., 1997; Van de Koppel, Prins, 1998; Makarieva, Gorshkov, 2007), Fig. 1.

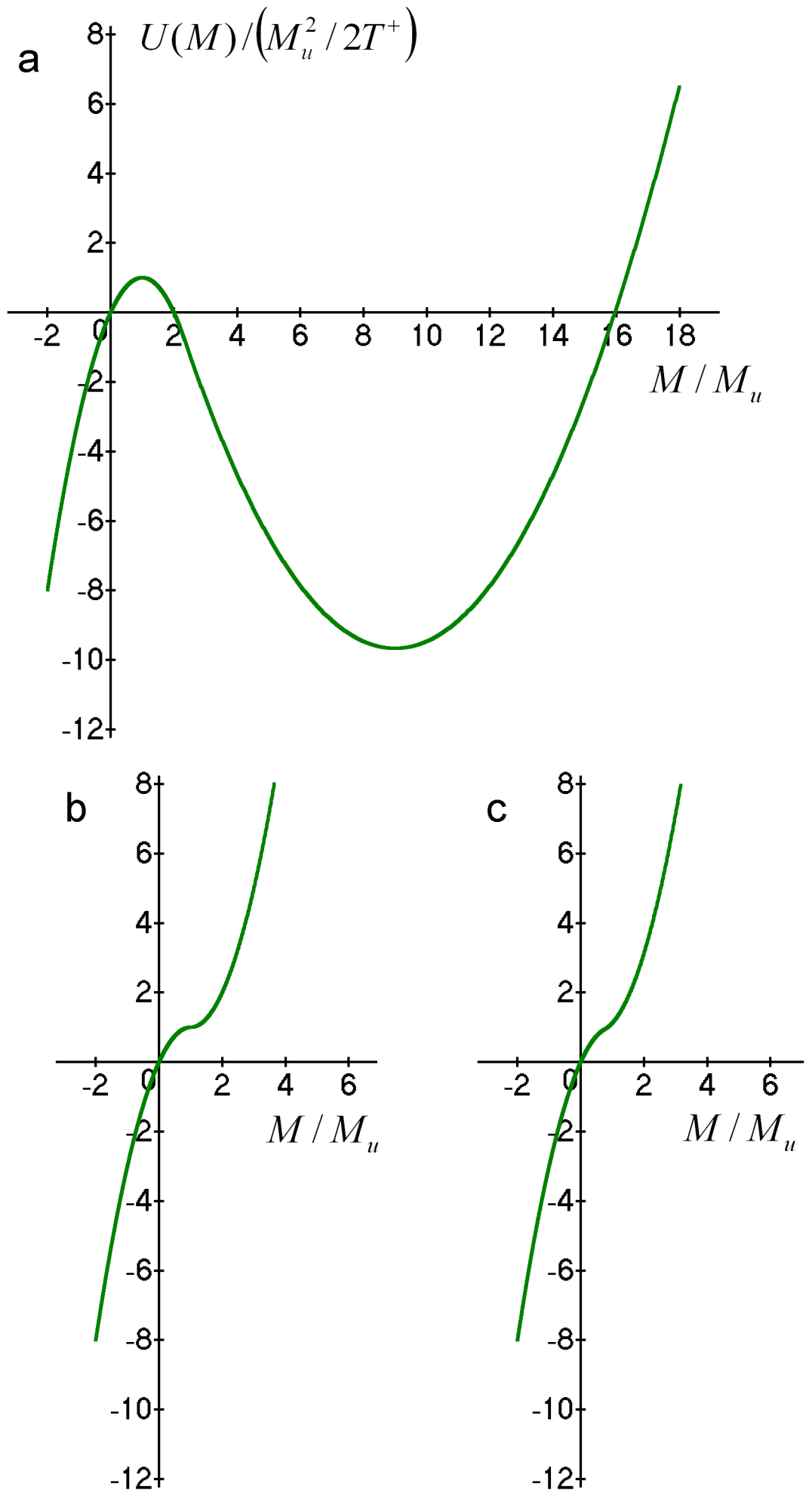

Fig. 1. Potential Lyapunov function $U(M)$ (6) for ecosystem biomass $M$ for different values of parameters of Eq. (6), which correspond to different states of the ecosystem. Units of the vertical and horizontal axes are dimensionless. At non-existent negative values of $M$ function $U(M)$ is shown for the purpose of generality. See text for further discussion.

(legend to Fig. 1, continued)

**(a)** Stable natural forest and unstable grassland (low index $a$ in parameters of Eq. (6)). Forest production in terms of organic carbon mass is $P_a^+ = 8$ t C (ha · year)$^{-1}$, time of forest biomass turnover via its decomposition is $T_a^- = 19$ years, forest biomass is $M_{sa} = P_a^+ T_a^- = 150$ t C · ha$^{-1}$; grassland decomposition is $P_a^- = 4$ t C (ha · year)$^{-1}$, time of grassland biomass turnover via its production is $T_a^+ = 4$ years, grassland biomass $M_{ua} = P_a^- T_a^+ = 17$ t C · ha$^{-1}$; $\alpha_a = M_{ua} / M_{sa} = 0.11$ (numerical data from (Whittaker, Likens, 1975)).

**(b)** Instability of exploited ecosystems (low index $b$ in parameters of Eq. (6)).
1) Exploited forest periodically cut: time-averaged mean forest biomass (from clear cutting to recovery) is half the stationary biomass of natural undisturbed forest, $M_{sb} = M_{ub} = M_{sa}/2 = 70$ t C · ha$^{-1}$, decomposition and synthesis are equal to each other, $P_b^- = P_b^+ = P_a^+ = 8$ t C (ha · year)$^{-1}$; turnover times via production and decomposition are equal to each other as well, $T_b^+ = T_b^- = 9$ years (turnover time for wood, which is produced ten times more slowly than the value of $P^+$, is about 90 years), $\alpha_b = M_{ub} / M_{sb} = 1$.
2) Unstable agriculture on clear-cut areas: with growing season of 0.5 year, the time-averaged (from sawing to yield) biomass is half the final biomass, which gives $M_{sb} = M_{ub} \approx M_{ua}/4 = 4$ t C · ha$^{-1}$, $P_b^- = P_b^+ = P_a^+ = 4$ t C (ha · year)$^{-1}$; $T_b^+ = T_b^- = 1$ year; $\alpha_b = M_{ub} / M_{sb} = 1$.

**(c)** Non-stationarity of biomass after introduction of alien herbivores, e.g., into an island ecosystem (low index $b$ in parameters of Eq. (6)): $P_c^- = 5$ t C (ha · year)$^{-1}$, initial biomass $M_{uc} = M_{ua} \approx 16$ t C · ha$^{-1}$, $P_c^- = 2 P_a^- = 2 P_a^+ = 8$ t C (ha · year)$^{-1}$; $T_c^+ = T_c^- = 2$ years; $M_{sc} = 8$ t C · ha$^{-1}$, $\alpha_c = M_{uc} / M_{sc} = 2$.

We emphasize that there are no physical laws that would prohibit complete decomposition of any particular type of biomass, as well as of the entire biomass of the biosphere, by biomass consumers, given that maximal productivity is limited from above by the externally fixed flux of solar radiation. In other words, the condition $\alpha << 1$ does not follow from any physical laws. Therefore, this condition should be written in the genetic information of the natural consumers in the biosphere (see, e.g., Makarieva et al., 2005). That this genetic information is not eroded in the course of accumulation of spontaneous mutations is ensured by competitive interaction of a large number (population) of local internally correlated communities of aboriginal biomass producers and consumers. When the genetic information of consumers in a local community is partially lost (eroded, undergoes decay), its biomass loses stability. The community loses competitiveness and is replaced by a normal community well before its biomass is totally exterminated by its genetically defective consumers (Gorshkov et al., 2004).

A most important consequence of the genetic principle of maintaining biomass stability is the rigid correlation of the aboriginal species of producers and consumers in any natural community. Only evolutionarily settled aboriginal species are able to maintain stable biomass in the community. This means that introduction of alien species into the community inevitably reduces or undermines completely the stability of community's biomass, because a random coincidence of the genetic properties of the aboriginal and invasive species is improbable, see Fig. 1b,c.

## 3. Stability of employment levels and prices in the market

The amount $M$ of some goods available at a particular time in the market in economics obeys the same equations of matter conservation as biomass in the biosphere, (1), (2) and (3). In this case $P^{+}(M)$ and $P^{-}(M)$ refer then to the processes of production and consumption of goods, respectively. In the previous section we have analyzed the stability principles governing maintenance of biomass in the processes of its production and consumption in the biosphere. The physical laws of energy and matter conservation indicate that there is a confined stability corridor harbored by the ecosystem for prolonged time periods of its existence, as prescribed by the genetic information of the community species maintained by natural selection. Life of all organisms in the biosphere, rendered possible when their biomass is stable in the long term, represents a supra-organized process maintaining life-compatible natural environment on Earth (Gorshkov et al., 2004; Makarieva, Gorshkov, 2007; Li et al., 2008).

Similarly, stability of economics is essential for the normal functioning of human beings, whose active work is aimed at sustaining the stability of the artificial highly-organized "environment" of the civilization where they live. These directional, non-random activities of humans should be well-organized and sufficient in power to firmly bar the civilization from spontaneous degradation.

The rate of goods output $P^+(M)$ can be related to number $N$ of people producing these goods as

$$P^+(M) = a\,N, \qquad (8)$$

where $a$ is a parameter independent of $N$, which has the meaning of *per capita* labour productivity. We will call $N$ employment. Relation (8) makes it possible to change our formalism from the one based on mass $M$, which can vary greatly from one type of goods to another, to the one based on measuring labor time and the corresponding number, $N$, of laborers. It is indeed labor time that is offered in the market, is sold and purchased, and it is proportional to the number of people who work. The labor dimension can be universally applied to all types of values offered by the market, both goods and services, which, as in the case of values produced by culture workers, priests, scientists, professional sportsmen or office cleaners, do not have a mass dimension.

An essential point in which the so-far considered ecology of the biosphere and economics of the civilization differ, is the fact that in the biosphere the primary basis of all processes is biomass production $P^+$ (2) constrained by the available flux of solar energy. The remaining fundamental parameters, $P^-$, $T^+$ and $T^-$, are organismal characteristics evolutionarily tuned to ensure stability of biomass and life in the ecosystem. In contrast, in economics the turnover times $T^+$ and $T^-$ of goods in the market and the saturated consumption rate $P_s^-$ are dictated by needs of the current population and cannot be easily changed. The value of $P_s^-$ is a function of the number of consumers only, while $T^-$ is a characteristic of goods. It is only the rate of goods output $P^+(M)$, and, hence, the number of people producing these goods that respond most easily to the change of the amount of goods in the market. Notably, a similar situation takes place in ecosystems of the open ocean, where the controlling ecological process is consumption by zooplankton of the primary production of phytoplankton (Gorshkov et al., 2000, Chapter 5.8). Here phytoplankton features a negligible, and zooplankton a major, part of the ecosystem biomass. In the meantime, the so far discussed interpretation of equation parameters in Section 2 pertains to those ecosystems where the major part of ecosystem biomass is represented by the biomass of primary producers, like terrestrial ecosystems, coral reefs and coastal ecosystems, Sargasso sea, etc. As will be shown below, the mathematical description of all types of ecosystems and economics remains essentially the same, it is only the meaning of measurable parameters that changes. (The unstable state with low biomass for open ocean corresponds to reduction in biomass of zooplankton and nekton.)

When consumption of conventional goods is saturated, and the demand is matched by the offer, the amount of these goods in the market reaches saturating value $M = M_s$, where $P^+(M_s) = P_s^-$. In this state the output rate should increase at $M < M_s$ (deficit of goods) and decrease at $M > M_s$ (overproduction). Within the stationary point $M = M_s$, where consumption $P^-(M)$ remains saturated and does not depend on $M$, function $P^+(M)$, which satisfies the above conditions, has the form:

$$P^+(M) = P_s^-\left(2 - \frac{M}{M_s}\right), \quad P^-(M) = P_s^- . \tag{9}$$

Eq. (1) for the change of the amount of goods $M$ in the market coincides then with Eq. (2) at $M_s \equiv P^+(M_s)T^- = P_s^- T^-$.

When new products of technological progress first appear in the market, their original amount is small. Production rate grows more rapidly than consumption. Therefore, as in the case of low ecosystem biomass (3), one can put production proportional to goods amount keeping consumption constant:

$$P^+(M) = \frac{M}{T^+}, \quad P^-(M) = P_u^- . \tag{10}$$

In this case equation (1) coincides with Eq. (3) at $M_u = P_u^- T^+$. Stability of the amount of goods in the market is described by the same Lyapunov function (5), (6), (7), as is the stability of biomass in the biosphere.

Using definition (8) and relationships (9) and (10) one can formulate the following equations for employment $N$:

$$\frac{dN}{dt} = \frac{N_s - N}{T^-}, \quad N_s \equiv P_s^- / a = M_s /(aT^-) ; \tag{11}$$

$$\frac{dN}{dt} = \frac{N - N_u}{T^+}, \quad N_u \equiv P_u^- / a = M_u /(aT^+) . \tag{12}$$

Equations (11) and (12) and the corresponding Lyapunov function differ from equations (2) and (3) and Lyapunov function (6), (7) by replacement of $M$ by $N$, $M_s$ by $N_s$, $M_u$ by $N_u$ and $M_r$ by $N_r$, see (5). Equations (11) and (12) contain the employment variable $N$, turnover times $T^-$ and $T^+$ (which, for goods and services of the civilization, always practically coincide) and two parameters, $N_s$ and $N_u$, having the dimension of employment. Parameter $N_s$ has the meaning of the level of employment which saturates production and consumption of the conventional goods of the civilization in the stable equilibrium state, Fig. 1a. Parameter $N_u$ can be interpreted as the critical level of employment in the beginning of the production process. The stationary point $N = N_u$ is unstable, Fig. 1a. If the novel goods are welcomed and accepted by the consumers, the employment (and, hence, production and consumption of novel goods) grows up to the saturated stable value $N = N_s$, where the novel goods enter the category of conventional goods. If, for some reasons, production and consumption of the goods cannot be increased, the employment falls down to a threshold value dictated by the basic demand of the population (as in the case of oil and other energy sources, see the following sections) or down to zero when the goods vanish from the market altogether (unsuccessful novel goods).

In economics, to allow for effective exchange of different goods and services one uses the universal monetary dimension. In free market economy the monetary price of goods $D$ is linked to the amount of goods in the market; it rises when the goods are in deficit and falls when they are abundant. The simplest dependence of $D$ on $M$ can be written as

$$D \equiv \frac{C}{M}, \qquad (13)$$

where coefficient $C$ does not depend on $M$. Substituting (13) into (2) and (3) we obtain, respectively:

$$\frac{dD}{dt} = -\frac{D}{T^-}\left(1 - \frac{D}{D_s}\right), \quad D \le D_r, \quad D_s \equiv \frac{C}{M_s}; \qquad (14)$$

$$\frac{dD}{dt} = \frac{D}{T^+}\left(1 - \frac{D}{D_u}\right), \quad D \ge D_r, \quad D_u \equiv \frac{C}{M_u}. \qquad (15)$$

Equations (14) and (15) represent the classical non-linear Verhulst equations (see, e.g., Lotka, 1925). All the subsequent conclusions will stand with a more general relationship $D = f(C/M)$, cf. (13), where $f(x)$ is an arbitrary function monotonously increasing with $x$. Indeed, in the vicinity of the stationary points (14) and (15) $D_i$ ($i = s$ or $u$), $x - x_i << x_i$, we have $D = f[x_i + (x - x_i)] = f(x_i) + c(x - x_i)$, where $c$ is a constant. This corresponds to the equality $D - D_i = c(x - x_i)$ and is equivalent to Eq. (13). Only the first multiplier in (14) and (15), which determines the behavior of Lyapunov function $U(D)$ at $D \to 0$ and $M \to \infty$, and the particular character of the monotonous junctions between the stationary points may change, but these are insignificant for any conclusions.

Stability of market price of goods and services $D$ can be investigated with help of Lyapunov potential function $U$ defined as

$$\frac{dD}{dt} = -\frac{dU(D)}{dD} . \qquad (16)$$

The value of $D_r$ can be, similarly to (5), defined as the point where the derivatives $dU/dD$ (14) and (15) coincide:

$$D_r = \frac{C}{M_r} = \frac{(T^+ + T^-)D_s D_u}{(D_u T^+ + D_s T^-)} . \qquad (17)$$

Integrating (16) with use of Eqs. (14) and (15), setting one constant of integration equal to zero and choosing the other, $r_D$, from the condition of continuity of $U(D)$ at $D = D_r$, we obtain the following expression for Lyapunov function describing (14) and (15):

$$U(D) = \begin{cases} -\dfrac{D^2}{2T^-}\left(1 - \dfrac{2}{3}\dfrac{D}{D_s}\right), & D \le D_r, \\ \dfrac{D^2}{2T^+}\left(1 - \dfrac{2}{3}\dfrac{D}{D_u}\right) - r_D, & D \ge D_r, \end{cases} \qquad (18)$$

$$r_D \equiv D_r^2 \frac{T^+ + T^-}{6T^+T^-} . \qquad (19)$$

At $D_s/D_u \equiv N_u/N_s << 1$ ($D_s << D_u$), Lyapunov function (18) has two stable minima, $D = D_s$ and $D = D_t$, and two unstable maxima at $D = 0$ and $D = D_u$, Fig. 2. The first maximum at $D = 0$ corresponds to zero price of goods in the market and absence of economics. The first minimum corresponds to the minimum stationary price $D = D_s$, while the second unstable maximum corresponds to the maximum stationary price $D = D_u$. The second stable minimum $D = D_t$ arises due to the boundary condition, Fig. 2: the consumer, close to the threshold of economic competence, pays the maximum price for indispensable goods. A most important example of such goods sold at $D = D_t$ is energy and raw materials in the modern civilization, see Sections 5-8.

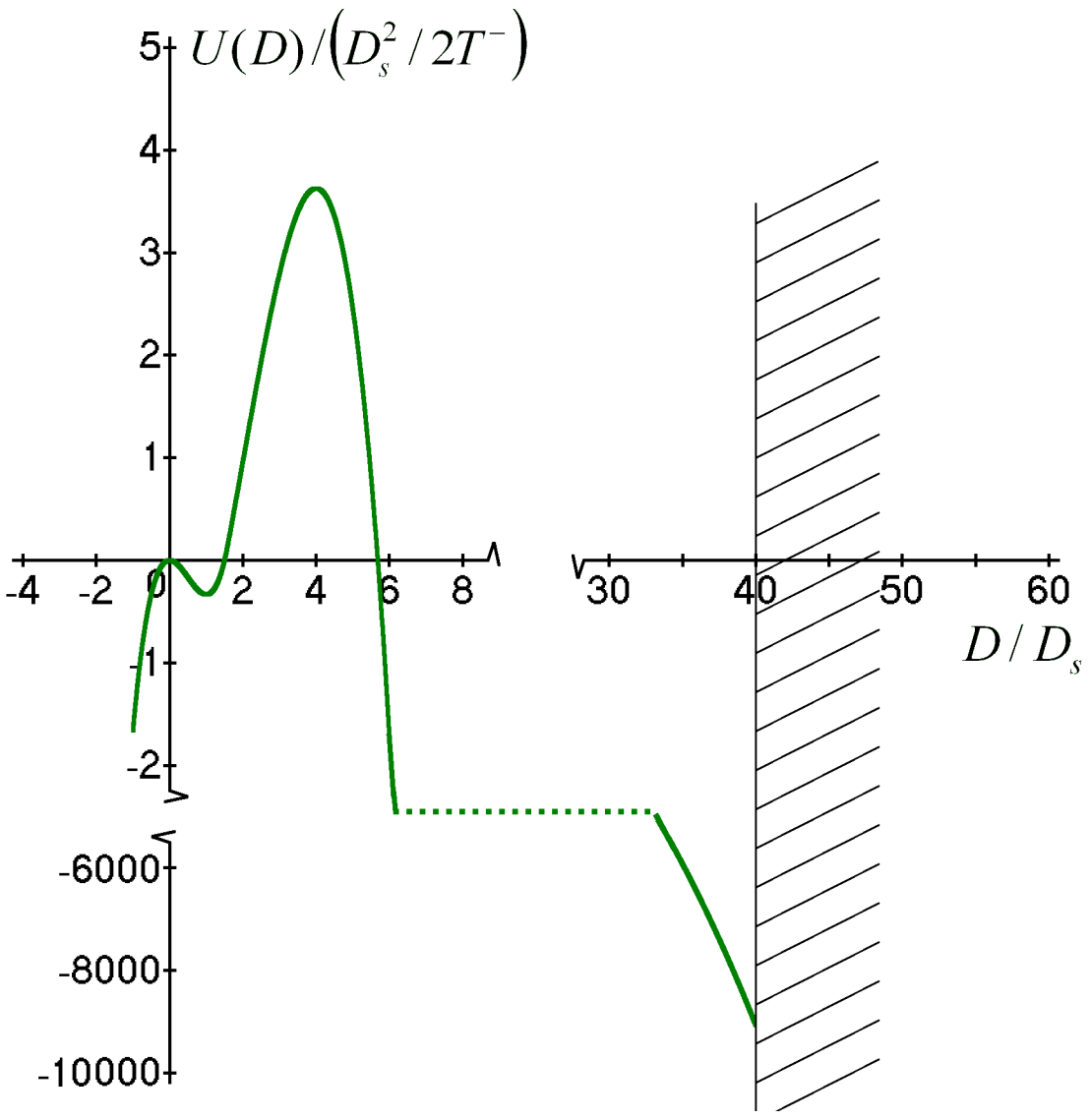

---

Fig. 2. Potential Lyapunov function $U(D)$ (18) for market price $D$ of goods in free market economy. Units of the vertical and horizontal axes are dimensionless. $D = D_s$ ($D/D_s = 1$) is the stable cost price of the conventional goods and services of the civilization ; $D_u \approx 4D_s$ is the unstable stationary price of novel expensive products of technological progress (Table 2); $D_t = 40D_s$ is the maximum stable price of raw materials and energy compatible with integrity of world economy (cumulative energy expenditures cannot exceed 10% GDP, see Fig. 3). The hatched area corresponds to breakdown of economics.

---

**4. Economic meaning of stationary market prices of goods**

All the numerous and diverse types of labor that are necessary to keep the civilization running are based on one and the same biological magnitude of food consumption by human body (metabolic rate). This metabolic rate is the universal energetic basis of all work performed by humans, be that control of technological devices or of social structures of humans themselves. This makes it possible to universally define the cost price of any type of goods or services as equal, to the accuracy of a constant multiplier, to labor time of people producing these goods or services. All types of goods and services produced, per unit working time, by the technologically armed workers of the civilization, will have one and the same cost price. Cost price paid to the worker by the consumer of the worker's production is sufficient to ensure normal living standard of the worker. Importantly, this "cost price" level of the living standard in the civilization is significantly higher than simply the level of survival (poverty threshold).

Labor time varies modestly from profession to profession, so it can be put equal to eight-hour working day for all activities. When one uses Eq. (8), all relationships for the market price of goods remain valid for the price of labor of the workers. In the result, the cost price of all goods and services delivered to the market can be considered proportional to the number of working people, see (8), with one and the same proportionality coefficient for all types of goods and services.

Cost price $D_s$ of any goods can be unambiguously determined, as discussed above, in the absolute terms of labor time. In terms of money, it can be defined as the price paid for this labor time to workers in the stable stationary state of potential minimum, Fig. 2a, $D = D_s$. Here the amount $M$ of goods in the market and number $N$ of people producing these goods have values $M_s$ and $N_s$, respectively, which correspond to saturated consumption $P_s^-$, see (8) and (13):

$$M_s = P^+T^-, \quad D_s = \frac{C}{M_s} = \frac{C}{P^+T^-}, \quad N_s = \frac{M_s}{aT^-} = \frac{P^+}{a}, \quad P^+ = P_s^- . \quad (20)$$

Using the known values $D_s$ and $N_s$ one can determine parameters $C$ and $a$ in (8) and (13):

$$C = D_s P_s^- T^-, \quad a = P_s^- / N_s . \quad (21)$$

As far as the overwhelming majority of workers produce conventional goods and services, while the proportion of workers producing novel expensive goods is negligible (see the sections to follow), it is possible to quantify the cost price via mean wages that on a global average are of the order of a few USD per hour (global GDP in 2005 being $45\times10^{12}$ USD, data of World Bank, www.worldbank.org, 2007; world population 6.3 billion people, assuming 50% working population, Table 1, and 2100 working hours per year, we obtain ~ 7 USD · hour$^{-1}$).

With the progress of civilization and change of technologies, there appear novel goods in the market that are attractive to the consumers. In the end of the XIXth such goods could be exemplified by the first ever telephones and automobiles, while towards the end of the XXth century those were mobile phones and personal computers. Until the market is saturated by such novel goods, the demand for them significantly exceeds the offer. Their market price and, hence, profit of the industry owner and potential wages of the workers, significantly exceed the cost price of the novel goods. The arising stationary point $D = D_u$, Fig. 2a, represents a critical value of the initial market price of the novel goods monopolistically produced by the pioneer manufacturers. This unstable price is critical with respect to either expansion or shrinkage of the output. High initial prices of the monopolistically produced novel goods and services ensure stable presence of such goods and services in the market.

Monopoly over production of novel expensive goods cannot be eternally sustained. Economic competition leads to an increase in the number of companies producing such goods and in the production rates, and makes the price of goods fall from some initial value $D$, $D_s < D < D_u$, towards cost price $D_s$. In the stationary state consumption and production become saturated at $P^- = P_s^-$ and $P^+ = P_s^-$. The goods are sold at their cost price $D_s$. Since then market price fluctuates stably around the stationary point $D = D_s$. When the competition in the sphere of goods output is well developed, the consumer is reluctant to buy the goods at a higher price $D > D_s$. Conversely, when the goods are overproduced, the market price falls below $D_s$, and the manufacturers have to reduce production, which leads to a rise of unemployment. Note that the greater the time $T^-$ of goods turnover in the market via their consumption (and, according to (8), turnover time of the corresponding labor), the slower and more painful this response to overproduction becomes. For example, for small goods like bolts and nuts with a short turnover time $T^-$, the overproduction is fixed practically instantaneously and cannot lead to massive lay-offs. In contrast, in automotive and aircraft industries or shipbuilding, where goods have inherently long turnover times, the overproduction becomes noticeable with a great delay, which leads to large numbers of workers getting fired.

With technological development of the civilization, one and the same number of workers becomes able to produce greater amounts of conventional goods per unit time. Consequently, the consumer can acquire greater amounts of conventional goods as payment for unit working time. This constitutes the essence of the civilization progress and is traditionally interpreted as increase in living standards. It is possible to sell novel expensive goods at prices significantly exceeding their cost price namely due to this important condition: the amount of money received by the worker when his production is sold at cost price is consistently greater than the poverty threshold, i.e. than the minimal income compatible with existence. People in the civilization can produce more goods than they need to simply remain alive. The difference between cost price level wages and minimal life-compatible income can be spent on buying the novel expensive goods.

The time of production of novel expensive goods is finite, from their initial appearance in the market to the moment of saturation of the demand with transition of the goods into the category of conventional goods, after which they are sold at cost price. During this time the owner of the production process acquires a finite sum of money. This sum arises due to the fact that some part of the population bought the novel expensive goods instead of some conventional goods. A certain amount of conventional goods thus remained unclaimed by the consumers. This effectively means overproduction of the conventional goods, which would not have occurred in the absence of novel expensive goods. The cumulative cost price of the conventional goods that remained unclaimed during the entire time of production of the novel expensive goods is equal to the sum of money received by the owner of the output of the novel goods. In the result, the unclaimed conventional goods are discarded and leave the economics, while the sum of money equal to their cost price, now owned by the producer of novel goods, remains available in the market. The excessive amounts of money compared to goods and services in the market leads to inflation, which is therefore an unavoidable economic by-product of the technological progress.

However, the possibility of making big fortunes by people involved in the production of novel goods stimulates further search of such goods and, thus, contributes to the technological progress of the civilization. This financial encouragement of economic and technological progress is widely recognized as an advantage of market economy. Significant technological advancements are relatively rare. The originally high prices of novel expensive goods fall rapidly down towards the cost price. At any given moment of time only a small share of goods in the market is being sold at prices greatly in excess of cost price. Therefore, inflation rates inherently associated technological progress should be relatively modest. The main contribution to inflation in modern economics comes from financial fluxes that arise when the civilization pay the high market prices for raw materials and energy, see Section 6.

## 5. Production of raw materials and energy in modern economics

All goods that are created by human labor are ultimately produced from raw materials with use of energy, both consumed from the environment. In modern economics raw materials and energy are considered as free market commodities, which, along with all goods produced by human labor, have a certain market price. This can be justified by the fact that the extraction – mining and quarrying – of raw materials and crude oil and other energy sources demands human labor. Thus, in accordance with Eqs. (15) and (16), raw materials and energy have a certain cost price.

The monetary value of cost price is, unlike its value in terms of labor time, numerically arbitrary and depends on the units of payments currently adopted in economics. In the imaginary case, when labor and production of all working people in the civilization is valued at cost price, the monetary contribution of every economic sector to the annual gross domestic product is proportional to the number of employees in this sector. Consider an economics consisting of two sectors with $N_1$ and $N_2$ employees. Let $G_s$ (dimension money units (e.g. dollars) per person per year) is the mean cost price of goods produced by one person in one year. If the goods made in the first sector of economics are sold at cost price, while in the second sector they are sold at a maximum price $G_t > G_s$ one can write

$$N_1 G_s + N_2 G_t = GDP\,, \tag{22}$$

where *GDP* is the annual gross domestic product. For the ratio of maximum and cost prices we obtain from (22)

$$\frac{G_t}{G_s} = \frac{N_1}{N_2} \frac{N_2 G_t}{GDP - N_2 G_t}\,. \tag{23}$$

With use of (23) it is possible to estimate how the current market price of oil is elevated above its cost price. Consider world economy as consisting of two sectors, where $N_2$ people are occupied with oil extraction and mining of other energy resources and related activities, while $N_1$ people produce all other goods and services of the civilization and sell them at cost price.

In 2005 world's nominal gross domestic product equaled $GDP = 45\times10^{12}$ US dollars (World Bank, 2007). This includes the price paid by the civilization for energy. Global energy consumption in 2005 amounted to $15\times10^{12}$ W = $4.7\times10^{11}$ GJ · year$^{-1}$ (see Appendix). Mean crude oil price in 2005 was 55 USD · barrel$^{-1}$, oil energy content is 5.5 GJ · barrel$^{-1}$, i.e. market price of 1 GJ was about 10 USD. Thus, for the price of energy produced and consumed in 2005 and for the term $N_2 G_t$ in (23) we obtain $N_2 G_t$= $4.7\times10^{11}\times10$ USD · year$^{-1}$ ≈ $5\times10^{12}$ USD · year$^{-1}$. Share of

**Table 1. Energy consumption, energy production, population and employment in 28 countries, including world's largest energy consumers and producers (2005)**

| Country | $P^-$, $10^{15}$ btu · year$^{-1}$ | $P^+$, $10^{15}$ btu · year$^{-1}$ | $N_{tot}$, $10^3$ | $N$, $10^3$ | $N/N_{tot}$, % |
|---|---|---|---|---|---|
| U.S.A. | 100.6 | 69.6 | 295374 | 141730 | 48 |
| China (2002) | 67.1 | 63.2 | 1284500 | 737400 | 57 |
| Russia | 30.3 | 52.7 | 142754 | 68169 | 48 |
| Japan | 22.6 | 4.1 | 127660 | 63560 | 50 |
| India (1991) | 16.2 | 11.7 | 838568 | 314131 | 37 |
| Germany | 14.5 | 5.3 | 82465 | 36566 | 44 |
| Canada | 14.3 | 19.1 | 32805 | 16170 | 49 |
| France | 11.4 | 5.1 | 62947 | 24919 | 40 |
| United Kingdom | 10.0 | 8.7 | 58653 | 28166 | 48 |
| Brazil (2004) | 9.3 | 7.7 | 149760 | 84596 | 56 |
| Italy | 8.1 | 1.2 | 58135 | 22563 | 39 |
| Iran | 7.3 | 13.0 | 67032 | 19760 | 29 |
| Mexico | 6.9 | 10.3 | 103831 | 40792 | 39 |
| Saudi Arabia (2006) | 6.7 | 25.5 | 27003 | 7523 | 28 |
| Spain | 6.6 | 1.4 | 43141 | 18973 | 44 |
| Australia | 5.5 | 11.2 | 20329 | 9739 | 48 |
| Indonesia | 5.4 | 9.3 | 228896 | 94948 | 41 |
| South Africa | 5.0 | 6.1 | 44344 | 12301 | 28 |
| Venezuela (2006) | 3.1 | 8.2 | 26952 | 11225 | 42 |
| Argentina | 2.9 | 3.7 | 23405 | 9634 | 41 |
| Pakistan | 2.3 | 1.6 | 158782 | 42816 | 27 |
| United Arab Emirates (2000) | 2.3 | 7.6 | 4087 | 1779 | 44 |
| Norway | 2.1 | 10.7 | 4593 | 2289 | 50 |
| Algeria (2004) | 1.4 | 7.7 | 32100 | 7798 | 24 |
| Kuwait (1997) | 1.2 | 6.1 | 2209 | 1218 | 55 |
| Iraq (1997) | 1.2 | 4.1 | 20700 | 4862 | 23 |
| Nigeria (1986) | 1.1 | 6.5 | 98937 | 30776 | 31 |
| Libya (1973) | 0.8 | 4.0 | 2249 | 531 | 24 |
| **28 countries** | **366.2** | **385.4** | **4042211** | **1854934** | **46** |
| World | 462.8 | 460.1 | 6300000 | | |

**Notes to Table 1.**

**Notations.** $P^-$ and $P^+$ are energy consumption and energy production in 2005 ($10^{15}$ btu · year$^{-1}$), respectively; $N_{tot}$ is total population of the country (thousands of people), $N$ is total number of working people (thousands of people); employment levels in different economic sectors is given as percentage of the total employment $N$, **Food** – agriculture, forestry and fishing, **Mining** – mining and quarrying, this sector includes production of raw materials and non-renewable energy, **Manuf.** – manufacturing, **Electr.** – electricity, gas

**Table 1. (continued)**

| Country | Employment by economic activity, % $N$ | | | | | | | |
|---|---|---|---|---|---|---|---|---|
| | Food | Mining | Manuf. | Electro. | Constr. | Trade | Transp. | Other |
| U.S.A. | 1.6 | 0.4 | 11.5 | 0.8 | 7.9 | 21.7 | 4.4 | 51.8 |
| China | 44.1 | 0.8 | 11.3 | 0.4 | 5.3 | 6.7 | 2.8 | 28.7 |
| Russia | 10.2 | 1.8 | 18.4 | 2.9 | 6.7 | 17.1 | 9.2 | 33.8 |
| Japan | 4.4 | 0.0 | 18.4 | 0.6 | 8.9 | 24.1 | 6.1 | 37.5 |
| India | 61.0 | 0.6 | 9.4 | 0.4 | 1.8 | 5.6 | 2.7 | 18.6 |
| Germany | 2.4 | 0.3 | 22.0 | 0.9 | 6.6 | 17.9 | 5.3 | 44.7 |
| Canada | 2.7 | 1.3 | 13.6 | 0.8 | 6.3 | 23.8 | 7.1 | 44.3 |
| France | 3.8 | 0.2 | 16.6 | 0.8 | 6.8 | 16.8 | 6.4 | 48.7 |
| UK | 1.4 | 0.4 | 13.2 | 0.6 | 7.8 | 19.6 | 6.9 | 50.2 |
| Brazil | 21.0 | 0.4 | 13.9 | 0.4 | 6.3 | 20.9 | 4.6 | 32.6 |
| Italy | 4.2 | 0.2 | 21.4 | 0.7 | 8.5 | 19.8 | 5.5 | 39.7 |
| Iran | 24.9 | 0.6 | 18.4 | 1.0 | 10.4 | 15.1 | 8.8 | 20.8 |
| Mexico | 14.9 | 0.5 | 16.9 | 0.5 | 7.8 | 28.9 | 4.5 | 26.1 |
| Saudi Arabia | 4.0 | 1.4 | 6.7 | 1.1 | 11.1 | 19.3 | 3.9 | 52.7 |
| Spain | 5.3 | 0.3 | 16.4 | 0.6 | 12.4 | 22.0 | 5.9 | 37.1 |
| Australia | 3.9 | 1.0 | 11.4 | 0.9 | 9.1 | 26.2 | 6.7 | 40.8 |
| Indonesia | 44.0 | 0.9 | 12.3 | 0.2 | 4.7 | 19.9 | 5.8 | 12.2 |
| South Africa | 7.5 | 3.3 | 13.9 | 0.8 | 7.6 | 24.6 | 5.0 | 37.2 |
| Venezuela | 9.1 | 0.7 | 12.0 | 0.4 | 9.4 | 23.3 | 8.1 | 36.9 |
| Argentina | 1.1 | 0.3 | 14.1 | 0.5 | 8.5 | 23.5 | 6.7 | 45.2 |
| Pakistan | 43.0 | 0.1 | 13.7 | 0.7 | 5.8 | 14.8 | 5.7 | 16.1 |
| UAE | 7.9 | 2.3 | 11.0 | 1.0 | 19.1 | 17.4 | 7.1 | 34.2 |
| Norway | 3.3 | 1.5 | 11.6 | 0.7 | 6.9 | 18.4 | 6.6 | 50.9 |
| Algeria | 20.7 | 1.7 | 10.9 | 1.0 | 12.4 | 17.2 | 5.6 | 30.5 |
| Kuwait | 2.0 | 0.7 | 6.6 | 0.7 | 10.9 | 16.2 | 3.4 | 59.7 |
| Iraq | 19.5 | 0.6 | 4.5 | 0.6 | 4.5 | 18.6 | 5.6 | 46.0 |
| Nigeria | 43.1 | 0.0 | 4.1 | 0.4 | 1.8 | 24.1 | 3.6 | 22.9 |
| Libya | 23.0 | 2.3 | 4.1 | 1.9 | 16.8 | 7.2 | 8.3 | 36.6 |
| **28 countries** World | **34.6** | **0.7** | **12.3** | **0.6** | **5.5** | **12.6** | **4.1** | **29.8** |

**Notes to Table 1 (continued)**

and water supply, **Constr.** – construction, **Trade** – wholesale and retail trade, repair, hotels and restaurants, **Transp.** – transport, storage and communications, **Other** – other activities, including state employees, military forces and social sector. Numbers in the eight right columns sum up to 100%. Cumulative energy consumption and population of the 28 countries considered constitute 80% and 60% of the global totals, respectively. Countries are ranked in the order of decreasing energy consumption.

**Data sources.** Population size and employment rate data correspond to year 2005 unless a different year is stated in the first column (e.g., for China these data are for 2002), data are taken from the International Labor Office

**Notes to Table 1 (continued)**

Bureau of Statistics (http://laborsta.ilo.org), yearly statistics, Tables 1A, 1B, 2B. In cases when population size data were missing at http://laborsta.ilo.org, they were taken from the U.S. Census Bureau (http://www.census.gov/ipc/www/idb/summaries.html). Energy production/consumption data for 2005 ware taken from the U.S. Energy Information Agency (http://www.eia.doe.gov), 1 btu (British Thermal Unit) = 1055 J.

**Employment rates in the food production sector** (Food) differ greatly for two groups of studied countries. The first group includes 17 countries with total population of 0.5 billion people mean employment rate in this sector is 4.4% (range 1.1-10.2%). In the second group of 11 countries (India, China, Brazil, Mexico, Indonesia, Iran, Pakistan, Algeria, Iraq, Nigeria, Libya) employment rate varies from 15% to 61% with a mean of 33%. According to the formulae of Section 5, in countries from the second group food is sold at cost price. In countries from the first group food is sold at market prices significantly exceeding the cost price. The difference between market and cost prices remains within the distributive network.

---

world's working population employed in mining and quarrying (i.e. in the output of energy and raw materials) is negligibly small, Table 1, at about several tenths of per cent (0.7% for the selected countries). Assuming that approximately one half of them is engaged in the output of energy, $N_1/N_2 \approx 300$, and putting the available estimates of $N_1/N_2$, $N_2G_t$ and $GDP$ into (23) we conclude that the market price of oil is elevated above its cost price by about forty times, $G_t/G_s \approx 40$ and $D_t \approx 40D_s$. A similar calculation performed on the basis of the data of Table 1 for electric power, which makes up about 10% of total energy production, indicates that for electricity its market price is elevated above its cost price by approximately four times, $D_t \approx 4D_s$.

Privatization of the production process, which, as discussed in the previous section, contributes in some ways to the technological progress of the civilization, is widely practiced in the energy and raw materials sectors of economics. However, there are fundamental differences between energy and raw materials commodities and novel expensive goods of the civilization, Table 2. Consumption of energy and raw materials is saturated at their maximum market price. Market mechanisms that can lead to a radical fall in prices of novel goods when their output grows do not work

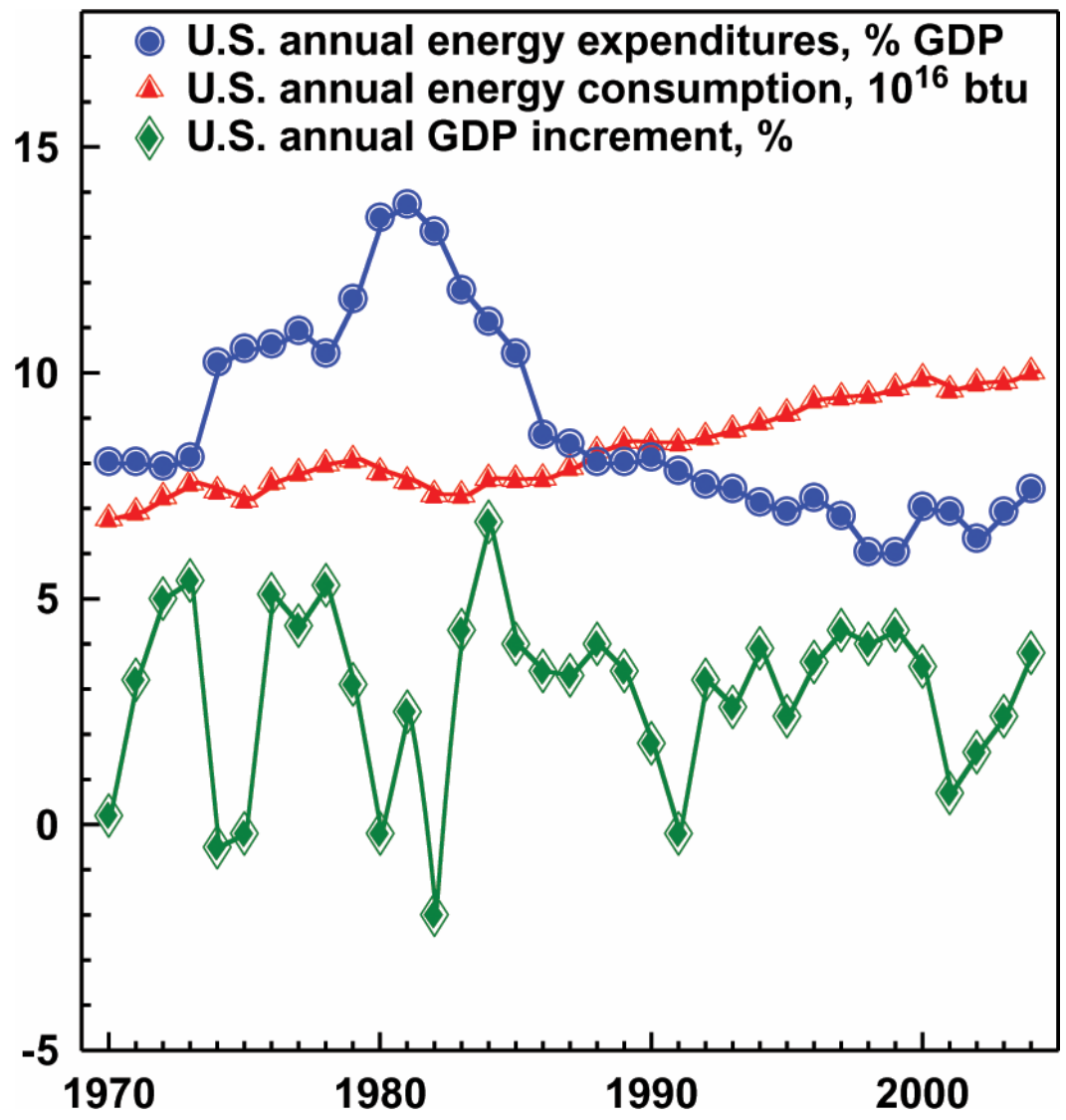

Fig. 3. Annual energy expenditures, energy consumption and GDP increment in U.S.A. in 1970-2004. Mean energy expenditures are 7% GDP. During the energy crisis of the 1980s, the twofold rise of energy expenditures resulted in economic instability when annual GDP increment dropped to the lowest negative values recorded during the entire period. When the energy expenditures returned to mean values, annual GDP increment and economics as a whole stabilized. This indicates that current energy expenditures (7% GDP in U.S.A., ~10% GDP on a global scale) represent the permissible threshold, beyond which economics starts to disintegrate. Data taken from the U.S. Energy Information Agency ( http://www.eia.doe.gov/emeu/aer/txt/ptb0105.html), 1 btu (British Thermal Unit) = 1055 J.

---

for energy and raw materials. The main idea behind free market is the idea of maximum profit for the business owner. Let us imagine one oil well owner increases the output and sells his oil at a lower price. As far as market price of oil exceeds its cost price by dozens of times, see above, all the other oil well owners are also able to easily cut their oil prices by a similar amount without reducing the output. The excessive oil amounts produced by the first oil well owner will remain unclaimed and fall out from economics with zero price. This owner will suffer losses, he will have to give up the idea of increasing the output and will have to elevate oil price back to the maximum. This consideration proves that free market competition in the sphere of production of raw materials and energy, the civilization essentials, is impossible.

Within the free market, owners of oil wells, ore mines etc. effectively remain monopolists on a global scale irrespective of whether they form a single correlated system, or represent multiple independent parties, as well as irrespective of whether they are individual businessmen within one country or particular countries within the world economy.

## 6. Quantifying inefficiency of an economy where energy is a market commodity

The civilization cannot exist without energy consumption. Therefore, the demand for energy is saturated irrespective of its market price, which, as discussed above, is kept at its maximum possible value. Energy consumers, i.e. the entire world economy, pay for energy with a certain share of the globally produced goods and services. As discussed in the previous section, these expenditures constitute about 10% of the annual gross domestic product. That this magnitude does not change consistently with time, Fig. 3, indicates that is already close to the stability threshold of the modern world economy, see Fig. 2.

Occupation within the energy production sector constitutes but tenths of per cent of the working population of Earth, Table 1. This means that, whatever luxuriant their living standards can be, people in this sector are unable to utilize the 10% of the globally produced goods and services paid to them in return for the energy they produce, Fig. 4. Therefore, these 10% of world's production must either be discarded as garbage and thus cause a 10% inflation rate, or used to support some population capable of their utilization. The size of this population supported at the existing mean living standard of the civilization (accounting also for maintenance of the elderly and the young) is of the order of the same ten per cent of global population, Fig. 4. This population, supported by the energy sector, does not necessarily have to participate in any meaningful activities stabilizing the civilization, for example they do not have to take part in industrial production. For this reason it is natural to call such population "vacant". We emphasize that maintenance of a vacant population is not an optional choice of the energy

**Table 2. Novel products of technological progress (NPTP) versus energy and raw materials (ERM): Comparison of their standings within a market economy**

| N | NPTP | ERM |
|---|---|---|
| 1 | NPTP are needed to, and consumed by, a certain part of population. | ERM are necessary to, and consumed by, all people in the civilization. |
| 2 | Expansion of output makes the original maximum market price of NPTP fall towards their cost price. The demand is saturated at cost price and the NPTP become conventional goods. | The demand for ERM is always saturated at a maximum market price that the civilization can afford to pay up to the threshold of losing economic integrity. |
| 3 | Market competition of independent parties producing NPTP, which facilitates expansion of output, is possible. | Market competition of independent parties producing ERM is impossible. ERM output is of effectively monopolistic nature. |
| 4 | Maximum price is only stable until the output starts to increase. | Maximum price is always stable; demand saturated. |
| 5 | Total profit obtained from the moment of NPTP appearance in the market, when they are sold at maximum price, to the moment when the demand is saturated at cost price, is a finite sum of money (dimension \$). | Profit from selling ERM under conditions of the demand saturated at a maximum price significantly exceeding cost price represents an infinite flux of money (dimension $\$ \cdot \text{year}^{-1}$). |
| 6 | The profit from selling NPTP is used to facilitate search of other NPTP, which contributes to the technological progress of the civilization and ensures economic growth even in a non-growing or shrinking population. | The profit from selling ERM is not used in the production process of the civilization. It is spent to support a growing part of the population, whose paid activities can take arbitrary forms (sports, religion, politics, terrorism, etc.). |
| 7 | The share of NPTP in the global consumer basket is negligible compared to the conventional goods. | Civilization spends over 10% of global annual gross domestic product to cope with the market prices of ERM that are elevated above the cost prices by dozens of times. |
| 8 | Privatization of the man-made output facilities (plants, factories) for producing NPTP does not violate fundamental physical laws. | Privatization of the natural sources for production of energy and raw materials (oil wells, ore mines) violates the fundamental physical law of conserved dimensions of measurable variables. |

sector, but the only opportunity to somehow utilize the enormous energy payments.

One can therefore envisage world economy as consisting of three sectors, which here can be, for the sake of definitiveness, referred to as the industrial sector (where all goods and services sustaining the current level of civilization development are produced), the energy production sector, and the energy-sponsored sector (vacant population supported by the energy production sector), Fig. 4. The financial mechanism of supporting the vacant population is as follows. Owners of energy production receive money after they sell energy in the free market. This money is then given to the vacant population so that they can buy in the same free market goods and services up to ten per cent of global production, which will otherwise remain unclaimed. Vacant population can be provided with a salary equal to the mean salary of the working population within the industrial sector.

Human being is genetically programmed to be active and cannot escape activity. Therefore, the vacant population is necessarily involved in some activities. These paid activities can take arbitrary forms, including various types of Sisyphean toil. Guided by wishes of the sponsors, these activities can be directed towards practically any goal, like creation of wonders of the world similar to Egyptian pyramids, organization of massive religious gatherings and sports competitions, implementation of political changes, as well as towards undermining the foundations of the society via, e.g., terrorism. These activities can also be directed towards stimulation of birth rate increase up to the female biological capacity and maximization of population growth rate. However, an overly high population numbers of vacant population, in particular, in excess of 10% of worlds' working population, will make their living standards drop below the global average. All paid activities of the vacant population are counted in the gross domestic product of the civilization, with the total contribution of the order of the same 10%, Fig. 4.

Social structure of the vacant population can be very diverse and complex. Vacant population can represent the majority of population in a particular country, or it can form several social or professional layers within a country where the majority of population is occupied within the industrial sector. Even a single individual can simultaneously represent both working and vacant population, e.g. working population who receive interest on the profit of companies drilling oil in their region. Importantly, to persist within the industrial sector, any activity must be economically competitive. Within the energy-sponsored sector the situation is fundamentally different – any activity must be in the first place liked by the sponsor, not necessarily economically competitive. Co-existence of these two types of activities within a society mixes up the societal values, misguides the economically active population and, as a result, slows down economic growth and reduces living standards.

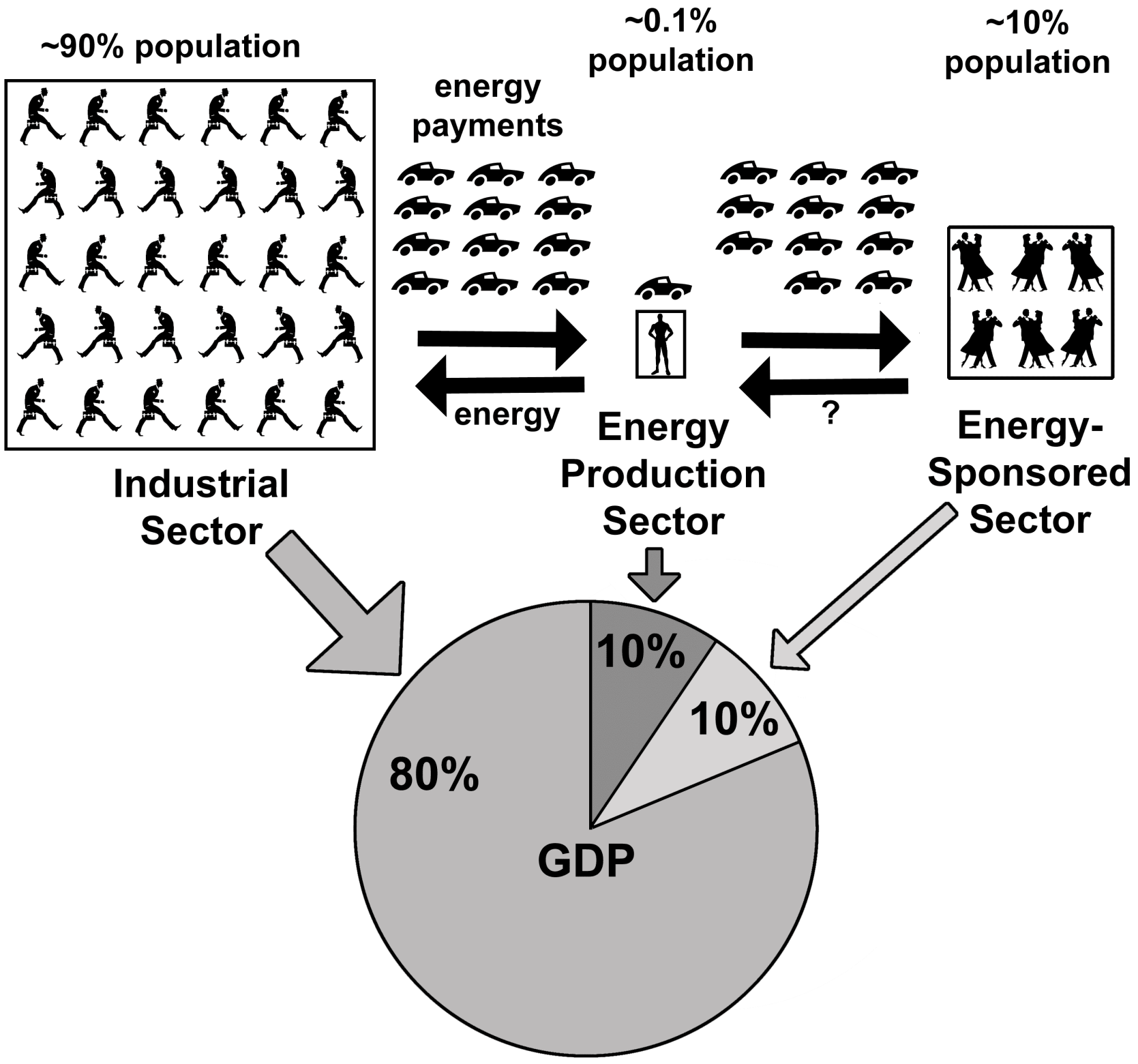

Fig. 4. Schematic representation of modern world economy with energy treated as a free market commodity. It is assumed for simplicity that the market price of energy is one hundred times higher than its cost price. So 0.1% population working in the energy production sector is paid 10% global GDP. This money is spent to support vacant population, which does not necessarily participate in the maintenance of civilization stability.

Prolonged existence in the state of vacancy of an entire population supported by the energy sector is equivalent to exclusion from active participation in the scientific and technological progress of the civilization. It results in decline of educational standards, disappearance of qualified specialists and general social degradation.

Vacant population is to be affected most when the cheap energy sources are exhausted. When the cost price of energy reaches up its current market price, the producers of energy will be unable to support a vacant population. The third, energy-supported, sector of economy vanishes, Fig. 5. Occupation in the energy production sector will have to rise from the current tenths of per cent to 10% of the working population. This is formally equivalent to vacant population getting engaged in energy production. Although the magnitude of GDP will formally fall by 10% with disappearance of paid arbitrary activities of the vacant population, neither the level of employment, nor living standards in the industrial sector will change. With energy cost price rising even further beyond current market prices, the share of people working in the energy production sector will have to rise as well. In the result, occupation in the industrial sector will start to fall making the scientific and technological level of the civilization, as well as the living standards of the population, decline.

In modern economy cheap energy resources still abound. If one abandoned considering energy as a free market commodity and energy were sold it at cost price worldwide, there would be no need to support vacant population and no need to produce the additional 10% of global goods and services for this purpose, Fig. 6. This would open a possibility of globally reducing working hours by ten per cent. This could be done, for example, by diminishing the five-day working week by half a day or reducing the retirement age by four years (given an average forty years working lifetime), all this without affecting the current living standards.

To summarize, considering energy as a free market commodity make the majority of working population of Earth spend ten per cent of their working time to support a vacant population, whose activities are not necessarily contributing to the stability of the civilization and are not necessarily economically competitive.

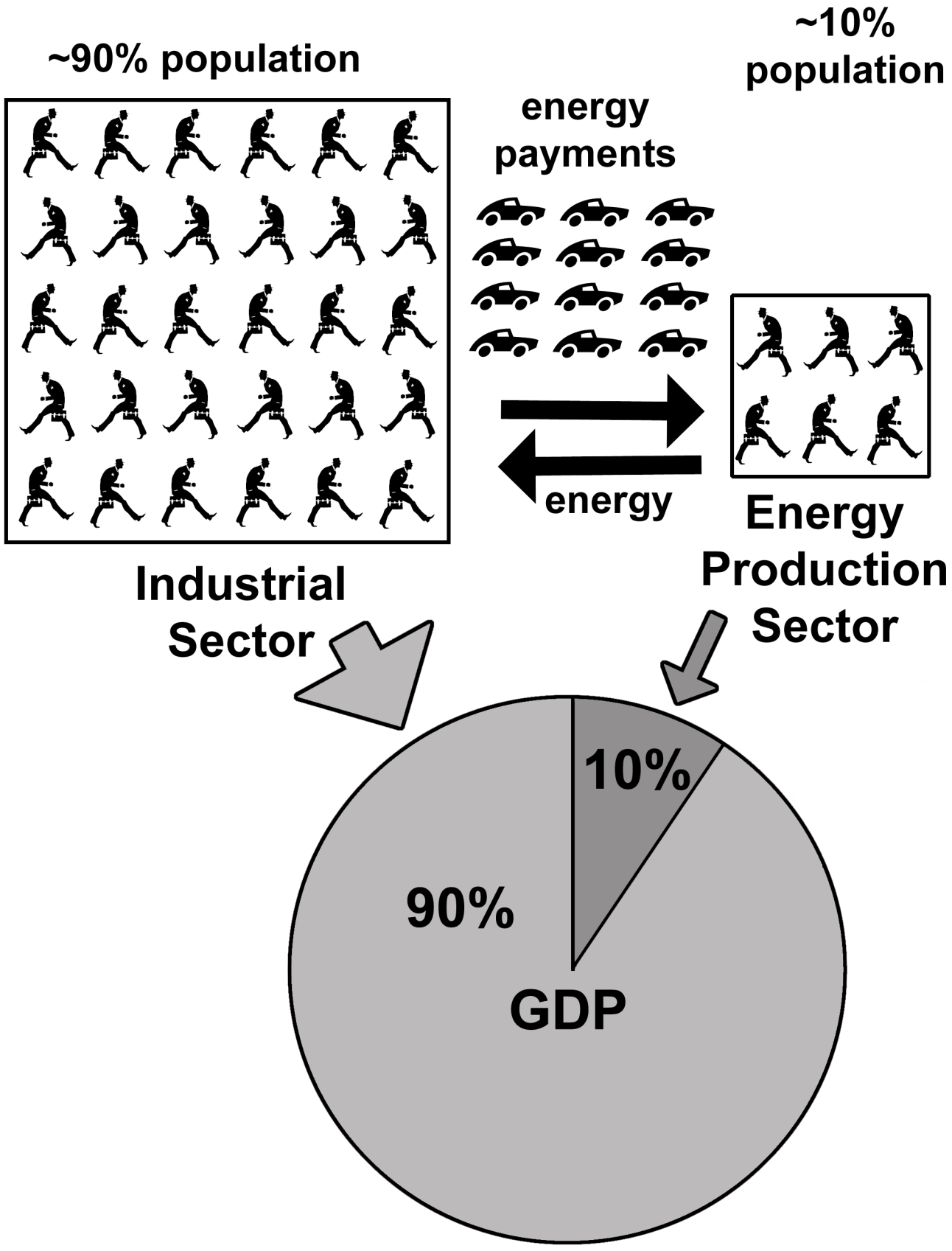

Fig. 5. Schematic representation of world economy based on expensive energy sources. It is assumed that all cheap energy sources are exhausted. Energy has a high cost price coinciding with its modern market price considered in Fig. 4. Maintenance of vacant population is no longer affordable, cf. Fig. 4. Effectively, vacant population moves to energy sector.

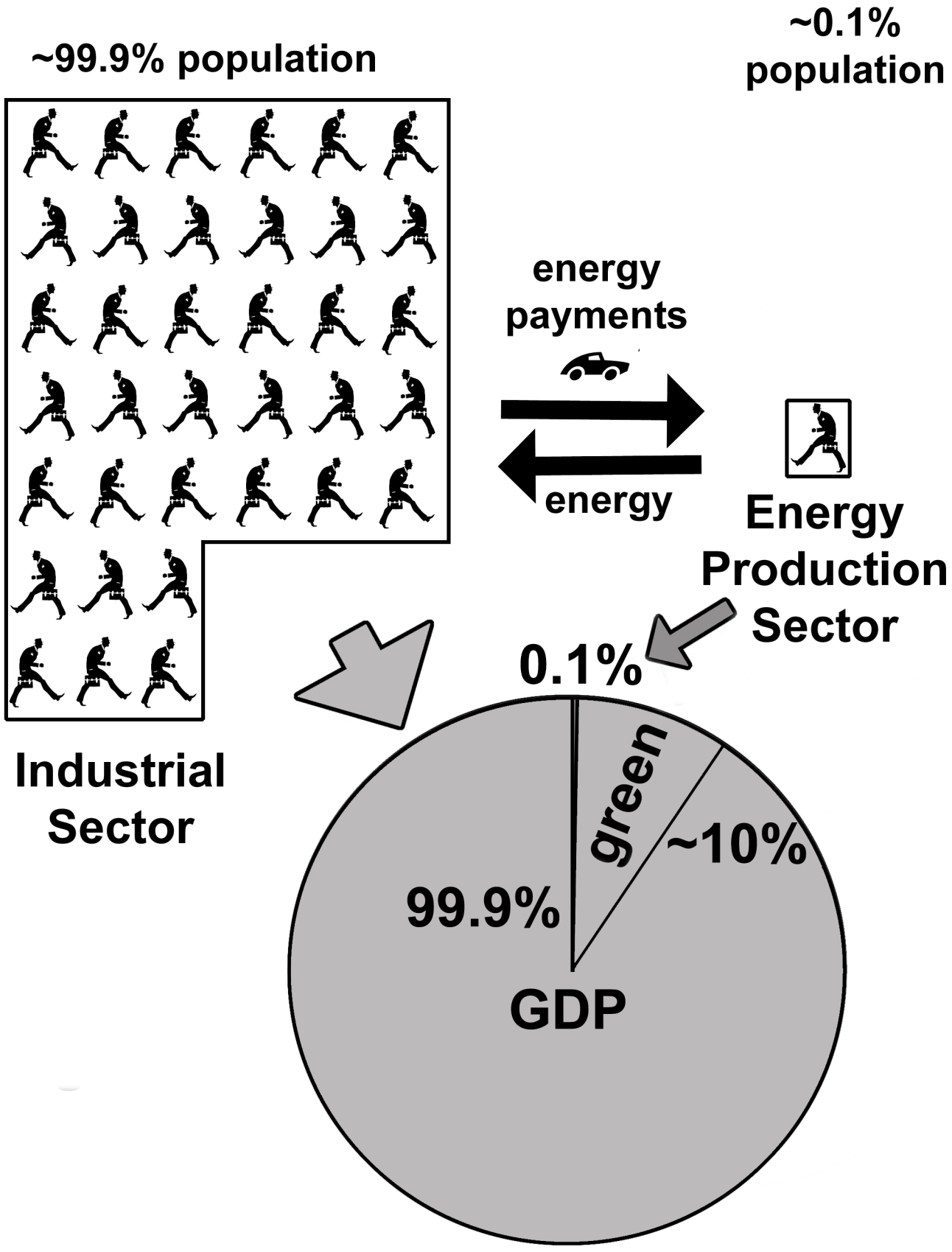

Fig. 6. Schematic representation of world economy based on cheap energy sources, which are legally excluded from the list of market commodities and are sold at cost price. Market price of energy is reduced down to its current cost price (cf. Fig. 5 where, conversely, cost price of energy is raised up to current market price). Effectively, vacant population moves to the industrial sector. In the result, there appears a green share of global GDP of the order of 10%, which can be used for solution of environmental problems without reducing modern living standards.

## 7. Ownership of energy sources as a physical error of economic theory

Modern economic theory is based on the notion of ownership. Economic laws prohibit forceful exemption of property from the owner. The owner has the right to ask any price for his property and sell his property for a finite sum of money. We will now show that the existing state of world economy, when the civilization pays forty times the cost price of energy on the verge of losing economic integrity, is conditioned by the incorrect definition of ownership in the economic theory, which recognizes the ownership right over sources of raw materials and energy like working hydropower stations, oil wells, ore and coal mines, as well as over natural production units as agricultural lands and forests.

All ordered processes in the biosphere and civilization are based on energy consumption. Biomass can be measured in the units of mass $M$, as well as in the units of energy using mean energy content of organic matter, which for live biomass approximates 4.2 $kJ \cdot g^{-1}$. Similarly, any article produced by the civilization can be quantified in mass units, as well as in the units of labor, i.e. in terms of work on its production, see (15), (16). Human work is proportional to human metabolic rate (~100 W) multiplied by working time (hours, weeks or years) and, hence, has the dimension of energy (J). Pricing goods and services available in the market in monetary units essentially represents pricing the corresponding human work. Money, thus, represents a unit of measuring energy and can be converted to conventional energy units – Joules.

As shown in the previous sections, the cost price of energy is dozens of times lower than its market price. In the approximation of zero cost price of energy one can convert the real annual production of the civilization, which in monetary units is equal to $35\times10^{12}$ USD (global GDP minus energy expenditures minus activities of vacant population, see Section 6 and Fig. 4), into energy units using the known value of global energy consumption, $0.47\times10^{12}$ $GJ \cdot year^{-1}$. Equating these two magnitudes we obtain a conversion factor from dollars to Joules, 1 USD = 13 MJ, or 1 GJ = 75 USD. (These figures are not related to, and do not coincide with, the market price of 1 GJ of energy!) Money is thus an energy unit used to describe the production process of the civilization. In other words, dollars and Joules are different units of one and the same dimension, energy.

Money is also a mass unit for the goods that are being produced. However, very different amounts of different goods can be produced in one and the same time unit of human work. Monetary mass units for different goods can therefore differ by orders of magnitude; they are inconvenient and are not used.

Fluxes of energy and raw materials, as well as fluxes of food, have a different physical dimension, not equivalent to mass or energy. This is the dimension of units of mass or energy *per unit time*. These fluxes cannot therefore be quantified in units of money. For example, a hydropower plant is characterized by the electric power it generates. The dimension of power is Watt, i.e. Joules per second. The dimension of power cannot be transformed to the dimension of energy (which, as we showed, is economically equivalent to the monetary dimension) in very much the same manner as the dimension of time cannot be transformed to the dimension of length or mass (i.e. one hour cannot be measured in kilograms). It is therefore impossible to economically justify a finite sum of money for which the river power of the hydropower plant can be bought or sold. Privatization or selling the hydropower plant for any finite sum of money is the same economically unjustified as its forced expropriation. Such transactions, when they happen, are therefore essentially of non-economic nature.

All said about hydropower is also relevant for oil wells, coal mines and, generally, stores of any raw materials and energy resources used by the civilization. These stores are commonly referred to as non-renewable. However, the estimated time to exhaustion of these stores are usually much longer than the characteristic time scale of even long-term economic forecasts. Therefore, in all economic estimates such stores are effectively considered unlimited and represent sources of energy. Therefore, like hydropower, they do not have a monetary price, i.e. there is no economic basis for them to be owned.

In modern economics various subjects (from individuals to countries) can possess the ownership right (sovereignty, in case of countries) with regard to sources of energy and raw materials. The owners are therefore legally permitted to ask any price for energy and raw materials that the civilization is able to pay. This ownership right leads in modern economics to the violation of the fundamental laws of energy and matter conservation implied equating measurable variables of different dimensions (financial and energetic fluxes to financial and energetic stores). Indeed, the owner is able to sell the infinite production of the owned source of energy and raw materials at a market price. The money the owner thus receives does not correspond to any work performed, because the cost price of this production is negligibly small compared to the market price. In other words, there appears an infinite flux of money (which is equivalent to a flux of mass or energy) out of nothing.

If this fundamental physical error of the economic theory were corrected, and the juridical possibility of the ownership right over sources of energy and raw materials excluded, the energy producers would be selling not energy and raw materials to the civilization, but their labor, as employees. If these labor costs were overestimated by forty times, it would be possible to find different employees working at a fairer price. For this purpose one would need a centralized body controlling global energy sources and functioning as the employer. Notably, human history already saw precedents when the juridical norms impeding economic progress were abolished, for example, when the right for owning a human being (slavery) was legally prohibited. Slavery implied purchase and sale not of human work (which has the dimension of Joules and coincides with the dimension of money), but of human power (which has the dimension of power and cannot be expressed as a sum of money). Remarkably, this abolished economic practice violated the physical law of conserved dimensions of measurable variables in very much the same manner as currently do the purchase and sale of energy sources.

## 8. Conclusions

Free market economy is credited for facilitation of the economic and technological progress by encouraging and stimulating corresponding activities in the population. In this paper we have argued that this is only true for the economic sectors dealing with the novel products of technological progress, but not with the conventional life-supporting goods and services without which human life in the modern civilization is impossible. The fundamental differences between these economic realities, Table 2, are not recognized by the modern economic theory. It was shown that, with current market prices of energy exceeding the cost price of energy production by at least forty times (Section 5), the civilization pays for this shortcoming of economic theory by approximately one tenth of the global gross domestic product. The money is ultimately spent to support some part of adult population (termed here “vacant”) who do not participate in the production process of the civilization, Fig. 4.

The humanity is currently facing serious problems of regaining and conserving the stability of the human-friendly environment and climate on all scales. The developed countries are widely criticized for their unwillingness to undertake adequate financial efforts to solve these problems (e.g., Stern, 2006). However, such critiques miss the important issue that under the burden of dramatically overpaying for energy, the economies of developed countries already function on the verge of losing their integrity, see Figs. 2 and 3. Until the economic and legal status of energy within the world economy is re-analyzed and changed, the hopes for an active and real participation of developed countries in the global environmental efforts will remain ungrounded.

If energy were legally excluded from the list of free market commodities and sold at cost price, and with the transition of the vacant population to the production sector of the world economy, cf. Fig. 4 and Fig. 6, this would grant the humanity a potentially "green" tenth of global GDP, Fig. 6. This share of GDP could be used for solving the environmental problems **with no change in the living standards and without threatening the economic development of the civilization.** As a first measure to be taken, the green GDP tenth could be used to urgently stop deforestation in the tropical countries. As was recently shown (Makarieva et al., 2006; Makarieva, Gorshkov, 2007), natural forests are responsible for the persistence of the water cycle on land, as they represent a biotic pump which draws atmospheric moisture from ocean to land and compensates the gravitational river runoff. According to the data of Food and Agricultural Organization of the United Nations (FAO, 2007), approximately one half of annual global deforestation occurs in Brazil ($3\times10^6$ ha · year$^{-1}$) and Indonesia ($1.9\times10^6$ ha · year$^{-1}$), while gross domestic products of these countries make up about 2% of global GDP, i.e. only one fifth of its potentially "green" slice, Fig. 6.

**Acknowledgements.** A.M.M. and V.G.G. express their gratitude to Jos van Damme, whose question on possible ways out of the global environmental crisis catalyzed the ideas developed in this paper. The authors thank R.L. Kitching, A. Mitchell, W. Manfrinato, H. Murray-Philipson and M. Shahwahid for inspiring discussion of global economic and environmental issues in summer 2007. Work is partially supported by Global Canopy Program and Rainforest Concern (A.M.M. and V.G.G.) and Russian Fund for Basic Research (A.M.M.).

**APPENDIX. Quantifying the Energy Budget of the biosphere and civilization: An overview of global renewable and non-renewable resources from the environmental safety viewpoint**

## A1. Introduction

As human body cannot exist without food, the civilization, at every stage of its development, must consume energy at a certain rate. Modern civilization, with its global energy consumption rate of around 15 TW (1 TW = $10^{12}$ W), largely exists at the expense of fossil fuels (oil, gas and coal). Burning of fossil fuels leads to accumulation of carbon dioxide ($CO_2$) in the atmosphere.

From the second half of the $20^{th}$ century the so-called global change processes have been registered on the planet. These are manifested most unequivocally as the increasing frequency of regional climatic and biospheric anomalies of all kinds, including temperature extremes, fluctuations of the atmospheric and oceanic circulation and biological productivity etc. In parallel, it was found that the global concentration of atmospheric $CO_2$ (the second, after water vapor, most important greenhouse gas on Earth) is growing conspicuously, currently exceeding the preindustrial value by approximately 30%. These two observations were widely interpreted as unambiguously coupled by a cause-effect link ($CO_2$ accumulation as the cause, climate change as the effect). Accordingly, at the background of growing concerns about the state of the planet, the scientific and technological search for the so-called alternative (with zero or low $CO_2$ emissions) energy sources is steadily intensifying. (There is another, quite unrelated, reason for this search: the anticipated fossil fuel exhaustion.) The conceptual basis for such an approach to the energy/environment problem consists in the statement that the absence of direct anthropogenic pollution is the single —necessary and sufficient — condition for the environment to remain stable and human-friendly.

During the same period when the global climate changes started to be monitored, there were, apart from $CO_2$ accumulation, other global processes in action, with their decisive impact on climate and environmental stability remaining largely overlooked in the conventional paradigm (Gorshkov et al. 2002; 2004; Li et al. 2008). The conventional energy/environment paradigm does not take into account the degree to which the environment is controlled by the global biota, the latter developing power by several orders of magnitude larger than does the modern civilization. By the end of the 20th century the anthropogenic disturbance of the biota had amounted to over 60% of land area (World

Resources 1988) and the environmental controlling functioning of the biota was globally disrupted. We argue that namely this fact, so far undervalued in its importance, rather than direct anthropogenic pollution of the planet, is the primary cause of the global change. Accordingly, environmental stability can only be restored by reducing the anthropogenic pressure on the biota. This is impossible without reducing the global rate of energy consumption of the civilization.

Here we review the available, and perform several original, estimates of the major natural energy fluxes in the biosphere (Section A2). We further analyze how the energy use is structured in the modern civilization and how the energetic needs of the civilization should be re-organized to be met without compromising the global environmental safety (Section A3).

Note the following energy units, approximate relationships and constants that are useful for comparing numerical data from various data sources: 1 kWh $\cdot$ year$^{-1}$ = 0.1 W; 1 btu (British thermal unit) = 1 kJ; 1 g oil = 50 kJ; 1 (barrel oil) $\cdot$ day$^{-1}$ = 60 kW; $10^5$ btu $\cdot$ year$^{-1}$ = 3.3 W.

## A2. Energy budget of the biosphere

The main energy fluxes existing in the biosphere are estimated in Table A1.

### *A2.1. Energy of solar and thermal radiation*

All major physical and biological processes on the Earth's surface are supported by solar radiation. The power of solar energy flux reaching the planet outside the atmosphere is $1.7\times10^5$ TW (1 TW $\equiv 10^{12}$ W $\equiv 10^{12}$ J $\cdot$ s$^{-1}$). The ordered, spatially and temporarily concentrated fluxes of geothermal energy (geysers, volcanoes, earthquakes) are millions of times less powerful and, globally, do not exert any noticeable impact on the biotic and physicochemical processes (Table A1). The power of tides related to the Earth’s rotation around its axis is more than two hundred thousands of times less than the power of solar radiation; so tides are energetically globally negligible as well (Table A1).

About 30% of the solar radiation flux is reflected by the planet back to space, mostly by clouds. The remaining $1.2\times10^5$ TW of solar radiation flux is absorbed by the Earth’s surface and the atmosphere and is ultimately converted into thermal radiation. Thermal radiation leaving the Earth to space corresponds to a temperature of −18 °C. About 30% of solar radiation – approximately the same amount as is reflected into space, is absorbed by the atmosphere (again clouds mostly). Thus, it is around $8\times10^4$ TW of solar power that ultimately reaches the surface. This power supports all ordered physical and biological processes on the Earth's surface, including the civilization.

Flux of thermal radiation emitted by the Earth's surface to the atmosphere is equal to $2\times10^5$ TW, i.e. it exceeds the flux of solar radiation absorbed by the planet by 40%. This difference arises due to the so-called greenhouse effect of the atmosphere. The atmosphere (or, more precisely, its certain greenhouse components like water vapor, cloudiness and $CO_2$) plays the role of a planetary "coat", which returns 40% of surface thermal radiation back to the surface. This elevates the mean global surface temperature to +15 °C from −18 °C that would have been observed in the absence of greenhouse effect.

Temperature difference between the surface and thermal radiation emitted into space from the absorption bands of greenhouse gas molecules, is equal to 33 °C. This difference does not coincide with the temperature difference between the surface and the upper troposphere. As everybody knows from everyday's life, the latter difference is two-three times greater and is, therefore, unrelated to the magnitude of the planetary greenhouse effect (Makarieva and Gorshkov, 2007). (Greenhouse effect can exist at zero value of the vertical gradient of air temperature). The magnitude of the greenhouse effect is determined by the relative width of thermal spectrum that is covered by the absorption bands of the greenhouse substances. The absorption bands of water vapor and clouds cover practically the entire thermal spectrum, thus largely determining the magnitude of the planetary greenhouse effect. The absorption bands of $CO_2$ correspond to only 19% of the thermal spectrum width. Therefore, the increase of $CO_2$ in the atmosphere exerts only a minor influence on the greenhouse effect compared to water vapor, but it can considerably decrease air temperature in the upper atmosphere (Makarieva and Gorshkov 2007). The cumulative thermal radiation into space will remain to be largely determined by the absorption bands of water vapor and clouds and will correspond to the temperature difference of the same 33 °C between the surface and thermal radiation emitted into space. This fact is ignored in many studies (Makarieva et al. 2004) and leads to an incorrect evaluation of the impact of atmospheric $CO_2$ accumulation on global mean surface temperature.

*A2.2 Estimating the power of thermohaline circulation*

Thermohaline circulation is the global overturning of the ocean, with water masses sinking in the polar regions and upwelling elsewhere at lower latitudes.

**Table A1. Energy budget of the Earth's surface, 1 TW ≡$10^{12}$ W**

| Power | Nature | | Source |
|---|---|---|---|
| | ***Total Earth*** | | |
| Solar radiation | $8\times10^4$ | | *1* |
| Evaporation | $4\times10^4$ | | *1,2* |
| Sensible heat fluxes | $2\times10^4$ | | *1,3* |
| Thermohaline oceanic circulation | $10^3$ | | *4* |
| Atmospheric circulation (wind power) | $10^3$ | | *5* |
| Photosynthesis | $10^2$ | | *6* |
| | ***Land*** | Civili-zation | |
| Solar radiation | $3\times10^4$ | 0.004 | *1, [7]* |
| Evaporation | $5\times10^3$ | n.u. | *8,9* |
| Transpiration | $3\times10^3$ | n.u. | *8,9* |
| Atmospheric circulation (wind power) | 300 | 0.01 | *5, [7]* |
| Photosynthesis | 60 | 6 (40)[a] | *6, [10]* |
| River hydropower | 3 | 0.3 | *8, [11]* |
| Osmotic transition river-sea | 3 | n.u. | *12* |
| Oceanic waves | 3 | 0.0001[b] | *13, [7]* |
| Tides | 1 | 0.0001[b] | *14, [7]* |
| Geothermal (concentrated) | 0.3 | 0.01[c] | *15, [16]* |
| Anthropogenic energy consumption | | 15 | *[11]* |

**Notes to Table A1:**

[a]6 TW is direct consumption of primary productivity (food of people and cattle and wood consumption); ~40 TW is the photosynthetic power of the biota on territories disturbed by anthropogenic activities (~60% of land area).

[b]Oceanic waves and tides combined.

[c]Civilization can only use concentrated sources of geothermal energy, like, e.g., geysers; total geothermal power on Earth is of the order of 15-30 TW (Berman 1975; Hammons 2007).

n.u. not used

Sources of data (sources in square brackets refer to civilization): 1 – Ramanathan 1987; Schneider 1989; 2 – Makarieva and Gorshkov 2007; 3 – Palmen and Newton 1969; Makarieva and Gorshkov 2006; 4 – Section A2.2; 5 – Section A2.3; 6 – Whittaker and Likens 1975; 7 – International Energy Agency (www.iea.org) Statistics for Renewables in 2004; 8 – L'vovitch 1979; 9 – Brutsaert 1982; 10 – Gorshkov 1995; 11 – Energy Information Administration, Official Energy Statistics from the U.S. Government (www.eia.doe.gov), data for 2005; 12 – Section A2.4; 13 – Akulichev 2006; 14 – Hubbert 1971; 15 – Starr 1971; 16 – Hammons 2007.

Global mean temperature of the oceanic surface is 15 °C. Oceanic waters below 1 km have constant temperature of 4 °C world over. In the absence of thermohaline circulation, oceanic waters would have had uniform temperature at all depths. The reason for the constant low oceanic temperatures at depths below 1 km consists in the unique physical properties of water. Water has maximum density, i.e. it is the heaviest, at 4 °C. In the result, the cold polar waters sink to the depth of the ocean. The power of this downward flux, which occurs around the poles, is $F \approx 10^{15}$ $m^3 \cdot year^{-1} = 3\times10^7\ m^3 \cdot s^{-1}$ (Stuiver and Quay 1983). To compensate for this flux, water masses undergo upwelling over the remaining area of the world ocean, $S = 3.6\times10^{14}\ m^2$. The water masses ascend and heat up to the surface temperature. The mean upwelling velocity is $u = F/S \approx 2\ m \cdot year^{-1}$ $= 5\times10^{-8}\ m \cdot s^{-1}$. The waters are warmed from 4 °C to 15 °C, i.e. $\Delta T = 11$ K, during their upwelling from depth to the surface, at the expense of solar radiation. The energy flux of this heating is $\rho c \Delta T u$, where $\rho = 10^3\ kg \cdot m^{-3}$ is water density, $c = 4.2\ kJ \cdot (kg \cdot K)^{-1}$ is water heat capacity. The total global power of thermohaline circulation is thus $\rho c \Delta T u S = \rho c \Delta T F =$ $1.4\times10^{15}\ W \approx 1\times10^3$ TW, Table A1.

*A2.3 Estimating the power of atmospheric circulation*

Power of the wind atmospheric circulation is supported by solar energy. It creates upwelling vertical air fluxes that are compensated by the opposite horizontal flows of air at the surface and in the upper atmosphere. The upwelling water vapor fluxes lead to condensation of water vapor in the upper atmosphere and precipitation of the condensed moisture. In the result, the upwelling air fluxes move at the same velocity $w$. The global mean value of $w$ is determined by the known value of global precipitation $\overline{P} = 10^3$ kg $H_2O$ $m^{-2}$ $year^{-1}$ = 1.7 mmol $H_2O$ $m^{-2}$ $sec^{-1}$. At global mean surface temperature of 15 °C concentration of water vapor near the surface is $N_{H_2O} =$ 0.7mol $H_2O$ $m^{-3}$. Therefore the mean velocity of the upwelling water vapor and air is $\overline{E} / N_{H_2O} = 1$ mm $sec^{-1}$. The force making air masses rise appears due to the observed compression of the vertical distribution of water vapor, which is due to water vapor condensation in the upper atmosphere (Tverskoi 1951; Weaver and Ramanathan 1995; Makarieva et al. 2004; Makarieva et al. 2006; Makarieva and Gorshkov 2007). This force is equal to $f_E = \beta\gamma\rho g$, where $\beta \approx 3$ is the compression coefficient, $\gamma = 2\cdot10^{-2}$ is the relative content of saturated water vapor near the surface at global mean surface temperature of 15 °C, $\rho \approx 1\ kg \cdot m^{-3}$ is air density, $g = 9.8$ m · $s^{-2}$ is the acceleration of gravity. The power of upwelling air masses is equal to $f_E w h_{H_2O} S \sim 10^3$ TW, where $h_{H_2O}$ = 2.4 km is the scale height of water vapor vertical distribution in the atmosphere, $S = 5 \cdot 10^{14}$ $m^2$ is the Earth's surface area.

The power of the upwelling air fluxes is compensated by the dissipative power related to horizontal wind velocity *u*. This power is determined by surface friction force $f_T = \rho \nu \partial^2 u / \partial z^2 \approx \rho \nu u / l^2$ (Lorenz 1967; Landau and Lifshitz 1987), where $\nu \approx 3$ $m^2 \cdot s^{-1}$ is eddy viscosity (Fang and Tung, 1999; Makarieva and Gorshkov 2007); *z* is height; $l \sim 100$ m is the scale of the change of horizontal wind velocity over height, it describes the region where the friction force is non-zero. The power of dissipation is $f_T u l S \approx \rho u^2 l^{-1} S$. At the observed global mean horizontal wind velocity $u$ = = 7 m · $s^{-1}$ (Gustavson 1979) the dissipative power of global atmospheric circulation is about $10^3$ TW, Table A1, i.e. it coincides with the power of the upwelling air fluxes. The obtained strict physical estimate is based on the basic physical characteristics of the terrestrial atmosphere and coincides in its order of magnitude with the available phenomenological estimates (Gustavson 1979).

### *A2.4 Estimating hydropower and the power of the osmotic transition river-sea*

Gross global hydropower is estimated as the power of global river runoff, which is, on average, 300 mm · $year^{-1}$ over all land or $R = 1.5 \times 10^6$ $m^3 \cdot s^{-1}$, falling down from the mean height *H* of the continents, $H$ = 200 m. The resulting power is equal to $R \rho g H$ = 3 TW, where *g* is acceleration of gravity (L'vovitch 1979). The economically available hydropower is estimated as 20% of the gross physical power, i.e. at 0.6 TW (Asarin and Danilov-Daniljan 2006). At present the civilization already claims half of the economically available hydropower, Table A1.

Theoretical possibility of using power of the river-ocean transition is based on the salinity difference between river and oceanic water. One needs a semi-permeable membrane, which separates river and oceanic water and prevents the dissolved ions from penetrating to fresh water. In the meantime, the solvent (water) diffuses unimpeded through the membrane from the area of its higher concentration (fresh water) to the area of its lower concentration (oceanic water), which is the essence of osmosis. In the result, pressure and level of the oceanic water increases. The magnitude of this increase depends on the amount of dissolved ions, i.e. on the salinity of seawater.

The osmotic pressure of seawater dissolved ions (mean salinity 3.5%) equals 28 atmospheres. In terms of water column height, atmospheric pressure equals 10 m. Thus, the osmotic pressure of seawater can rise the level of seawater in a reservoir separated from river water by a semi-permeable membrane to up to 280 m. This is somewhat higher, but of the same order of magnitude, as the mean height of the continental river runoff. The theoretical power of using the osmotic transition river-ocean is therefore of the same order of magnitude as the global hydropower, Table A1. However, technical difficulties in creating the semi-permeable membranes firmly preclude the energetic use of the osmotic river-ocean transition.

**A3. Energy consumption and the environmental problems of our civilization**

World total energy production and consumption is about 15 TW = $1.5\times10^{13}$ W, Table A2. Energy consumption is partitioned between the non-renewable energy sources like oil (40%), coal (30%), gas (20%), and nuclear power (7%) and renewable energy sources, of which the most important is hydropower. It makes up 2% of total energy consumption, Table A2. All the other sources of renewable energy are negligibly small on a global scale (~1%).

Global production of electric power constitutes 11% of total energy production power. Deviations from this mean figure in particular countries are insignificant, Table A3. The main global energy source for production of electric power is fossil fuel burning (thermal electric power plants). Only in a few countries (France) a greater portion of electric power is produced from nuclear power.

Total energy consumption in France is equal to 400 GW, of which nuclear power accounts for 160 GW. It thus makes up 40% of total energy production, which is the largest relative amount in the world. Production/consumption of electric energy in France constitutes 55 GW, which is 14% of total energy consumption. Nuclear power stations produce 78% of electric energy, i.e. 43 GW or 11% of total energy consumption. This is less than one third (27%) of all power delivered by nuclear power stations. The remaining two thirds (73%) are spent on heat production. Predominantly, it is waste heat, because, due to its high radioactivity, it is very difficult to use it in livelihoods and industries. As is well-known, for safety reasons nuclear powers are built outside densely populated cities, which poses additional difficulty for utilizing nuclear heat. (The same ratio of usable to waste power is characteristic of all nuclear reactors functioning

**Table A2. Distribution of global energy consumption, TW (1 TW ≡ $10^{12}$ W), over energy sources in 2005.** Data of the U.S. Energy Information Agency (www.eia.doe.gov). The first four countries listed account for over half of the world's energy consumption.

| Region | Total | % Total | | | | | | |
|---|---|---|---|---|---|---|---|---|
| | TW | Oil | Coal | Gas | Nuclear | | Hydro | Other |
| **World** | **15** | **40** | **30** | **20** | **7** | **(2)** | **2** | **1** |
| USA | 3.4 | 40 | 22 | 30 | 7 | (2) | 1 | <1 |
| China | 2.3 | 20 | 75 | 3 | 1 | (0.3) | 1 | <1 |
| Russia | 1.0 | 20 | 20 | 55 | 3 | (1) | 2 | <1 |
| Japan | 0.75 | 53 | 20 | 14 | 12 | (3) | 1 | <1 |
| France | 0.40 | 40 | 4 | 15 | **40** | **(11)** | 1 | <1 |

**Notes to Table A2.** Civilian nuclear energy is exclusively used for generation of electricity. The efficiency of electric power production from nuclear power does not exceed 30%. Therefore, the usable share of nuclear power (figures in braces in Table A2) constitutes less than one third of total nuclear power production.

**Table A3. Distribution of electric power production over different energy sources (same data source as in Table A2, data for 2005)**

| Region | Total electric power, TW | Thermal (Oil, coal, gas) | Nucler | Hydro | Share of electric power in total energy consumption |
|---|---|---|---|---|---|
| **World** | 1.7 | 60% | 20% | 20% | **11%** |
| USA | 0.4 | 75% | 20% | 5% | **12%** |
| China | 0.2 | 85% | 2% | 15% | **9%** |
| Russia | 0.09 | 70% | 10% | 20% | **9%** |
| Japan | 0.09 | 60% | 30% | 10% | **12%** |
| France | 0.055 | 10% | **78%** | 12% | **14%** |

in Japan, USA and Russia). Therefore, consumption of usable energy in France is 400 GW – 160 GW + 43 GW ≈ 280 GW. Nuclear power contributes 15% to the consumption of usable energy in France, not 40% and by far not 80%, as one is periodically misinformed by the mass-media. The contribution of usable nuclear power into the total energy consumption of France is 11%, Table A3. Similarly, the global share of usable nuclear power in the total energy consumption of the world is only 2%, Table A2, which coincides with the contribution of hydropower.

To summarize, modern civilization practically exclusively relies on fossil fuels, which account for 90% of global energy consumption, Table A2. Fossil fuels are considered dangerous because of greenhouse gas ($CO_2$) emissions associated with their burning, hence the modern hunt for environmentally friendly renewables. Generally, there are two wide-spread, mutually exclusive attitutdes towards global climatic change. According to the first one, climate change does exist, it is of anthropogenic origin AND it is caused by $CO_2$ emissions. According to the second one, the existence of directional climatic change is reasonably questionable; but even if it does exist, it is not of anthropogenic origin and, in particular, is unrelated to $CO_2$ emissions. However, as already noted, although $CO_2$ is a greenhouse gas, it is not the major one in the atmosphere of Earth. All greenhouse substances — water vapor, cloudiness and $CO_2$ — are involved into biogeochemical cycles and controlled by the natural biota of Earth (Makar'eva and Gorshkov 2001; Makarieva and Gorshkov 2002). This control is mainly implemented through regulation of the atmospheric amounts of water vapor and clouds, which represent the major greenhouse substances of Earth's atmosphere. Therefore, namely the anthropogenic disturbance of this controlling mechanism, natural ecosystems, by deforestation on land in the first place, is responsible for the climate anomalies that are being observed with increasing frequency. Thus, we argue, breaking the conventional dichotomy of approaches to climate change, that, although climate change does exist and is of anthropogenic origin, its primary cause IS NOT the build-up of atmospheric $CO_2$.

The second widely appreciated disadvantage of using fossil fuels as the main energy source is their clearly foreseeable exhaustion. Easily accessible stores of oil, gas and uranium are to be depleted within a few coming decades, of coal — within one century. Here, again, search for alternative energy sources to replace the fossils appears to be an obvious strategy to pursue. It is commonly perceived as self-evident and hardly demanding any logical analysis. A fundamentally different strategy of solving the energetic and environmental problems of civilization follows, however, from the consideration of the biotic regulation mechanism. Let us quantify the amount by which the humanity has reduced the power of this mechanism. Total power of the global biota is of the order of 100 TW, of which land biota accounts for approximately two thirds. Humans have destroyed natural biota on 60% of land, which means the global regulatory biotic power has been reduced by approximately 40 TW. This figure comprises about 6 TW of the primary productivity of the biota consumed by the civilization in the form of food for people, cattle fodder and wood. The remaining power pertains to disturbed ecosystems incapable of performing environmental control.

The value of 40 TW characterizes the destabilising environmental impact of the humanity. It is determined by the scope of anthropogenic activities on the planet that are supported by direct energy consumption of 15 TW, Table A1. Thus, irrespective of whether anthropogenic energy consumption **is accompanied by environmental pollution (e.g., $CO_2$ emissions) or not,** as long as its magnitude remains unchanged, the destabilizing environmental impact of humans will persist as well. Notably, 40 TW is more than twice a higher power than the direct anthropogenic energy consumption. Global environmental stability can only be regained if one restores the biotic mechanism of climate control, which is possible if the energetic impact on the biota is reduced by at least one order of magnitude (Gorshkov et al. 2000).

Increasing energy efficiency is widely discussed as one of possible measures towards solution of both the energetic and environmental problems of the civilization. The destabilizing impact the humanity imposes on the biosphere is determined, however, by the useful power of energy consumption $P_u$, which fuels all anthropogenic activities. The efficiency of energy conversion, $\varepsilon$, is defined as the ratio of useful power $P_u$ to the total consumed power $P_t$, $\varepsilon \equiv P_u / P_t$ . If energy efficiency $\varepsilon$ is increased at fixed total power $P_t$, this means that the useful power available for anthropogenic transformation of the biosphere, increases as well. This will only stimulate further degradation of environmental stability. If energy efficiency $\varepsilon$ is increased at fixed useful power $P_u$, this means that the total consumed power $P_t$ decreases. This allows one to diminish the rate at which the resources are depleted and the environment is polluted by fossil fuel burning. However, this will not diminish the rate at which natural ecosystems are destroyed, because this rate is determined by the value of $P_u$. To decelarate degradation of the biotic regulation mechanism and to re-gain environmental stability one needs to reduce the global value of $P_u$, irrespective of whether energy efficiency is high or low. Therefore, although widely discussed in the environmental context, the problem of energy efficiency is logically unrelated to the problem of climate and environment stabilization.

Growth of energy consumption opens a possibility for acceleration of population growth up to the maximum possible rate determined by the biological reproductive capacity of the woman. In its turn, population growth demands increased energy consumption. Until fossi fuels are depleted, the power of energy consumption can grow arbitrarily rapidly, as dictated by the rate of demographic and economic growth. There is no logic in substituting fossil fuels by some pollution-free renewable energy sources. It is necessary to reduce the very magnitude of energy consumption and the related magnitude of the anthropogenic pressure on the biosphere. Growing population numbers, the associated growth of energy consumption and scope of human activities, degradation of the remaining natural ecosystems, can lead to an irreversible loss of climate and environmental stability well before the fossil fuels stores are depleted.

**A4. Discussion: The perspectives of using renewable energy sources**

Early in the history of the industrial burning of fossil fuels, their share in the global energy consumption was negligibly small. The environmental danger associated with fossil fuel burning started to be discussed only after its apparent consequences ($CO_2$ build-up in the atmosphere) became globally significant. At present the share of renewable energy sources in the energy consumption budget of the civilization is approximately as small as was the share of fossil fuels a century ago, Table A2. And again, as with respect to fossil fuels, there is a wide-spread and scientifically unverified opinion that the consumption rate of renewables can be safely increased to ultimately fully substitute fossil fuels, with global energy consumption rate remaining the same or even growing. This strategy is justified by the single environmental argument that the use of renewable energy sources is not accompanied by $CO_2$ emissions ("clean" energy sources). However, the available, although largely disregarded, scientific knowledge is already sufficient for this strategy to be rendered as directly facilitating environmental degradation.

Among the renewable energy sources currently used by the humanity, river hydropower is the most significant one, Table A2. However, all the available hydropower is one order of magnitude smaller than the modern global power of energy consumption, Table A1. At present, about half of the technologically available and economically relevant hydropower is already used up (L'vovitch 1979; Asarin and Danilov-Daniljan 2006). Use of hydropower by the civilization cannot be significantly increased.

The power of atmospheric circulation on land (wind power) is twenty times larger than the modern global power of energy consumption, Table A1. However, wind on land is controlled by the natural vegetative cover (forest) (Makarieva et al. 2006; Makarieva and Gorshkov 2007). Natural forests act as pumps of atmospheric moisture, with the forest-induced atmospheric circulation delivering moisture from the ocean to land. This process compensates the gravitational river runoff and supports soil moisture and biological productivity on land. Therefore, when the moisture-laiden ocean-to-land winds are, on their way to land, impeded by windmills, this steals moisture from the continent and undermines the water cycle on land; the more so, the greater the extent of the antropogenic consumption of wind power. In its effect, the use of wind power is equivalent to deforestation. Not to threaten the terrestrial water cycle, wind power stations can thus be allowed to exempt less than one per cent of the total wind power, which means no more than 5% of the modern rate of global energy consumption. Even this low figure is difficult to achieve due to the various technical limitations. To conclude, windmills will be never be able to compete in power with the existing hydrological dams.

Thermohaline circulation plays a most important role in the maintenance of the stability of the Earth's climate. The use of thermodynamic machine producing power at the expense of the difference between sea temperatures at the surface and at depths can dramatically aggravate the environmental problems of the humanity.

Solar power constitutes the basis of the energetic budget of the biosphere. Natural ecological communities of the biota maintain particular atmospheric concentrations of greenhouse substances with relatively narrow absorption bands but large optical depth (product of atmospheric height, absorption cross-section and concentration). Such greenhouse substances do not significantly influence the planetary greenhouse effect, see Section A2.1, but create a negative vertical temperature gradient in the terrestrial atmosphere. This brings about the observed non-equilibrium state of atmospheric water vapor, upwelling air fluxes and atmospheric circulation. Due to the large value of this gradient, 3/4 of solar power on the Earth's surface is transformed into the power of sensible and latent (evaporation) heat fluxes, Table A1 (Makarieva et al. 2006). On land, natural forest cover maintains optimal soil moistening and compensates river runoff by pumping atmospheric moisture evaporated from the ocean (Makarieva and Gorshkov 2007). Thus, photosynthesis performed by green plants consuming solar energy is the main controlling process for this life-supporting circulation (this process is already disrupted by human activities on two thirds of land area, Table A1).

In several billion years, life has evolved maximum efficiency in the use of solar energy for generation of all these life-supporting processes. Any large-scale consumption of solar energy, currently almost invariably perceived as harmless and environmentally friendly, will have a catastrophic impact on the resilience of these critically important life-supporting processes.

The main danger for the life-compatible environment and climate on Earth consists in the excessively high absolute magnitude of global anthropogenic energy consumption. This danger persists irrespective of whether this consumption is accompanied by direct environmental pollution like $CO_2$ emissions or not. Modern or, even more so, growing energy consumption will totally destroy Earth's environment within the coming decades. Within the nearest century it is necessary to reduce global energy consumption by one order of magnitude, when it can be met by the existing renewable energy sources without causing environmental and climatic problems. This demands a detailed scientific analysis of the biotic nature of environmental stability on Earth and of possible ways of how the imperative of reducing global energy consumption could be constructively faced. In the absence of such an analysis, the attempts to find alternative energy sources (including thermonuclear power) to sustain modern or growing energy consumption of the civilization represent a logical dead-end, which will result in but further aggravation of the dangerous situation the civilization today finds itself in.

**References to Appendix**

www.bioticregulation.ru/pubs/pubs2.php